\documentclass[onecolumn]{pasj01}

\usepackage{siunitx}
\usepackage[nooneline]{subfigure}
\subfiguretopcaptrue
\usepackage{ulem}
\usepackage{xcolor}
\usepackage[switch,mathlines]{lineno}
\usepackage{multirow}
\usepackage{booktabs}
\usepackage{tablefootnote}

\Received{2023 December 28}
\Accepted{2024 February 13}
\Published{2024 March 6}

\newcommand{\SGRB}{SGRB}

\newcommand{\SGRBs}{SGRBs}
\newcommand{\LGRBs}{LGRBs}
\newcommand{\swift}{\textit{Swift}}
\newcommand{\fermi}{\textit{Fermi}}
\newcommand{\integral}{\textit{INTEGRAL}}
\newcommand{\hete}{\textit{HETE-2}}

\newcommand{\Epeak}{\ensuremath{E_\mathrm{peak}}}
\newcommand{\thetaOBS}{\ensuremath{\theta_\mathrm{obs}}}
\newcommand{\thetaC}{\ensuremath{\theta_\mathrm{c}}}
\newcommand{\thetaCk}{\ensuremath{\theta_{\mathrm{c},k}}}

\newcommand{\thetaCO}{\ensuremath{\theta_\mathrm{c,0}}}

\newcommand{\fOk}{\ensuremath{f_{\mathrm{0},k}}}
\newcommand{\Eiso}{\ensuremath{E_\mathrm{iso}}}
\newcommand{\Liso}{\ensuremath{L_\mathrm{iso}}}

\newcommand{\Lobs}{\ensuremath{L_\mathrm{obs}}}
\newcommand{\Lmax}{\ensuremath{L_\mathrm{max}}}
\newcommand{\Lmaxk}{\ensuremath{L_{\mathrm{max},k}}}
\newcommand{\LmaxO}{\ensuremath{L_\mathrm{max,0}}}
\newcommand{\Llim}{\ensuremath{L_\mathrm{lim}}}
\newcommand{\Flim}{\ensuremath{F_\mathrm{lim}}}
\newcommand{\Nobs}{\ensuremath{N_\mathrm{obs}}}
\newcommand{\zobs}{\ensuremath{z_\mathrm{obs}}}

\newcommand{\Val}[3]{\ensuremath{{#1}_{-#2}^{+#3}}}
\newcommand{\VAL}[4]{\ensuremath{\left({#1}_{-#2}^{+#3}\right)\times{#4}}}

\usepackage{alphalph}

\begin{document}

\title{Peak energy--Isotropic Luminosity Correlation and Jet Opening Angle Evolution in {\swift}-BAT Short GRBs with Soft Tail Emission}

\author{Naoki \textsc{Ogino}\altaffilmark{1}$^{*}$}
\email{naokiogino@stu.kanazawa-u.ac.jp}
\author{Daisuke \textsc{Yonetoku}\altaffilmark{1}$^{*}$}
\email{yonetoku@astro.s.kanazawa-u.ac.jp}
\author{Makoto \textsc{Arimoto}\altaffilmark{1}}
\author{Tatsuya \textsc{Sawano}\altaffilmark{1}}
\author{Hamid \textsc{Hamidani}\altaffilmark{2}}

\altaffiltext{1}{College of Mathematics and Physics, School of Science and Engineering, Kanazawa University, Kakuma, Kanazawa, Ishikawa 920-1192, Japan}
\altaffiltext{2}{Astronomical Institute, Graduate School of Science, Tohoku University, Aoba, Sendai, Miyagi, 980-8578, Japan}


\maketitle


\begin{abstract}
   Some short gamma-ray bursts (SGRBs) exhibit a short duration and spectral hard emission (referred to as a ``hard spike'') followed by a slightly longer soft emission (known as a ``soft tail''). 
   We identified nine SGRBs with known redshift in the Swift/BAT gamma-ray burst catalog by specifically searching for the soft tail. 
   We found that spectra of these SGRBs can be described as a cutoff power-law model for both hard spike and soft tail, and both show time variation keeping the {\Epeak}--{\Liso} correlation. 
   This suggests that the emission mechanism of both phenomena is identical. 
   Furthermore, we found a trend of luminosity evolution as a function of redshift. 
   This phenomenon suggests that these bursts originate from sources that have intrinsically bright and/or energy density concentrated within a narrower jet at higher redshift. 
   We demonstrate that the average jet opening angle, derived from the jet break, can be explained by considering a model based on a strongly redshift-dependent jet opening angle. 
\end{abstract}


\section{Introduction}
Gamma-ray bursts (GRBs) are the most energetic electromagnetic explosions in the universe, releasing up to $10^{54}$\,erg in a short duration.
GRBs are classified into long GRBs ({\LGRBs}) with a duration of $>2$\,s and short GRBs (\SGRBs) with duration of $\le2$\,s.
{\LGRBs} and {\SGRBs} are thought to have different origins \citep{Kouveliotou+1993}.
{\LGRBs} are explained by the collapse model of massive stars \citep{Woosley1993,MacFadyen+1999} and have been confirmed by observations \citep{Iwamoto+1998,Hjorth+2003}.
On the other hand, {\SGRBs} are thought to originate from the merger of two compact objects, such as binary neutron star mergers or black hole-neutron star mergers \citep{Paczynski1986,Goodman1986,Eichler+1989}, but lacked observational evidence for a long time.

On August 17, 2017, the Advanced LIGO detectors \citep{LIGO+2015} and the Advanced Virgo detector \citep{Virgo+2015} observed gravitational wave GW~170817A, associated with the binary neutron star merger in the elliptical galaxy NGC~4993 \citep{Abbott+2017a,Abbott+2017b}.
Furthermore, $\sim1.7$\,s after the gravitational wave detection, the gamma-ray astronomy satellites {\fermi} \citep{McEnery+2012} and {\integral} \citep{Winkler+2003} simultaneously observed a weak $\gamma$-ray emission GRB~170817A, believed to be a {\SGRB} \citep{Abbott+2017c,Savchenko+2017}.
However, the properties of GRB~170817A are different from typical {\SGRBs}; it is known to have many orders of magnitude lower isotropic equivalent luminosity ({\Liso}) and isotropic equivalent energy ({\Eiso}) compared to other {\SGRBs} observed by {\fermi} \citep{Abbott+2017c}.
To explain the uniqueness of GRB~170817A, scenarios such as the off-axis jet scenario, where $\gamma$-rays are detected outside the beaming cone of a typical {\SGRB} jet \citep{Goldstein+2017,Murguia+2017,Granot+2018,Ioka+2018,Ioka+2019,Hamidani+2020}, and the cocoon shock-breakout scenario \citep{Kasliwal+2017,Bromberg+2018,Gottlieb+2018,Lazzati+2018,Nakar+2018,Hamidai+2023a,Hamidani+2023b} have been considered. 
However the mystery remains as GRB~170817A is the only event with gravitational wave and electromagnetic wave observation known so far.

Therefore, studies have been conducted to find GRB~170817A-like events or to examine the rate of such events among the {\SGRBs} that have been observed so far \citep{Mandhai+2018,Beniamini+2019,Kienlin+2019,Matsumoto+2020}.
GRB~170817A is known for its bright, hard, and short initial emission (hard spike) followed by a soft tail component (soft tail) shining at $\le50$\,keV.
In \citet{Kienlin+2019}, focusing on this property, a search for similar events in the {\fermi} gamma-ray burst monitor ({\fermi}-GBM) 10 yr burst catalog \citep{Kienlin+2020} was conducted, 
and ended up identifying 12 {\SGRBs} similar to GRB~170817A.
However, most of them have no redshift measurements, and the properties of rest frames such as {\Liso} and {\Eiso} are still unknown.
Hence, in this study we focused on {\SGRBs} observed by {\swift}/BAT with better localization than {\fermi}/GBM and selected events with the soft tail similar to GRB~170817A.

The structure of this paper is as follows.
In section \ref{sec:analysis}, we describe the method for event selection and present the results of the spectral analysis.
In section \ref{sec:correlation}, we investigate the {\Epeak}--{\Liso} correlation of the hard spike and soft tail, showing that all events exhibit similar temporal variations.
In section \ref{sec:discussion}, we deduce the structure of the jet (opening angle and viewing angle) from the {\Epeak}--{\Liso} correlation.
Finally, our concluding remarks are presented in section \ref{sec:conclusion}.

\section{Analysis} \label{sec:analysis}

\subsection{Event Selection using Light Curve} \label{subsec:lc}
The emission of GRB~170817A consists of a hard spike with photons in the 50--300\,keV energy range and a soft tail with photons $\le50$\,keV \citep{Goldstein+2017}.
\citet{Kienlin+2019} used bayesian block analysis \citep{Scargle1998,Scargle+2013} to define the hard spike and soft tail, selecting events similar to GRB~170817A from the {\SGRBs} observed by {\fermi}/GBM.
Here, following their approach, we applied bayesian block analysis to the light curves of {\SGRBs} observed by {\swift}/BAT to identify GRB~170817A-like events.

Initially, we selected 117 {\SGRBs} with a duration of $T_{90}\le2.0$\,s in the {\swift}/BAT Gamma-Ray Burst Catalog\footnote{https://swift.gsfc.nasa.gov/results/batgrbcat/} up to the end of April 2022.
Here, $T_{90}$ is defined as the time during which 90\% of the total observed counts are accumulated, excluding the first and last 5\%.
Then, as in \citet{Kienlin+2019}, we set the threshold of $50$\,keV and generated light curves in the 15--50\,keV and 50--150\,keV bands, defining the brightest time region in the 50--150\,keV as the hard spike.
If the bayesian block analysis identified a longer-lasting emission in the 15--50\,keV light curve beyond the time duration of hard spike in 50--150\,keV range, this emission was defined as the soft tail.
This method allowed us to select 43 events with the soft tail.
Note that bayesian block analysis was conducted using the \texttt{bayesian\_blocks} function of \texttt{astropy} version 5.2.2 \citep{Astropy2022}.
In this module, the parameter known as \texttt{ncp\_prior}, which influences the ease of detecting change points of bayesian blocks, was set to $\mathtt{ncp\_prior}=6.0$.
This value is standard in the analysis of {\swift}/BAT data \citep{Markwardt+2007}.

\begin{figure*}
   \centering
   \subfigure[GRB~060801]{
      \includegraphics[width=0.31\textwidth]{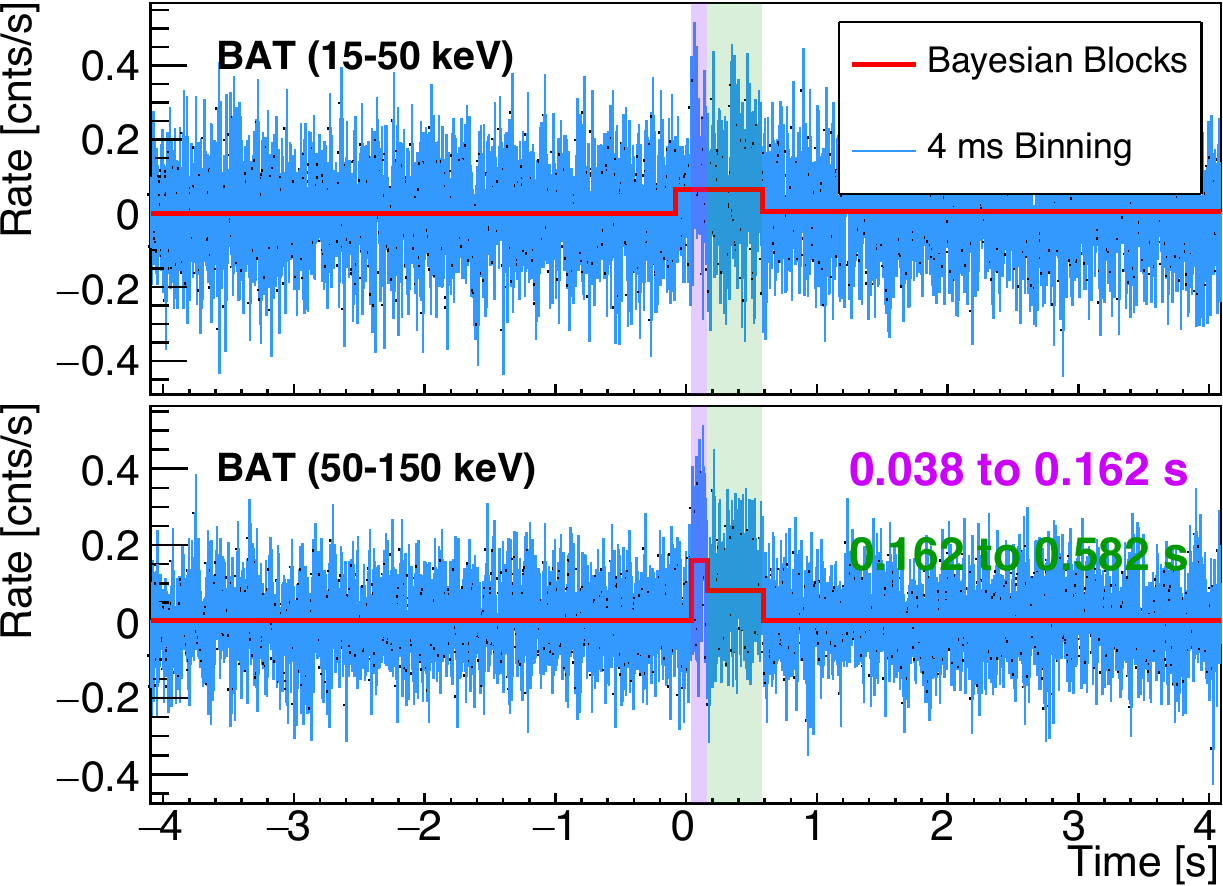}
      \label{fig:lc_060801}
   }
   \subfigure[GRB~100724A]{
      \includegraphics[width=0.31\textwidth]{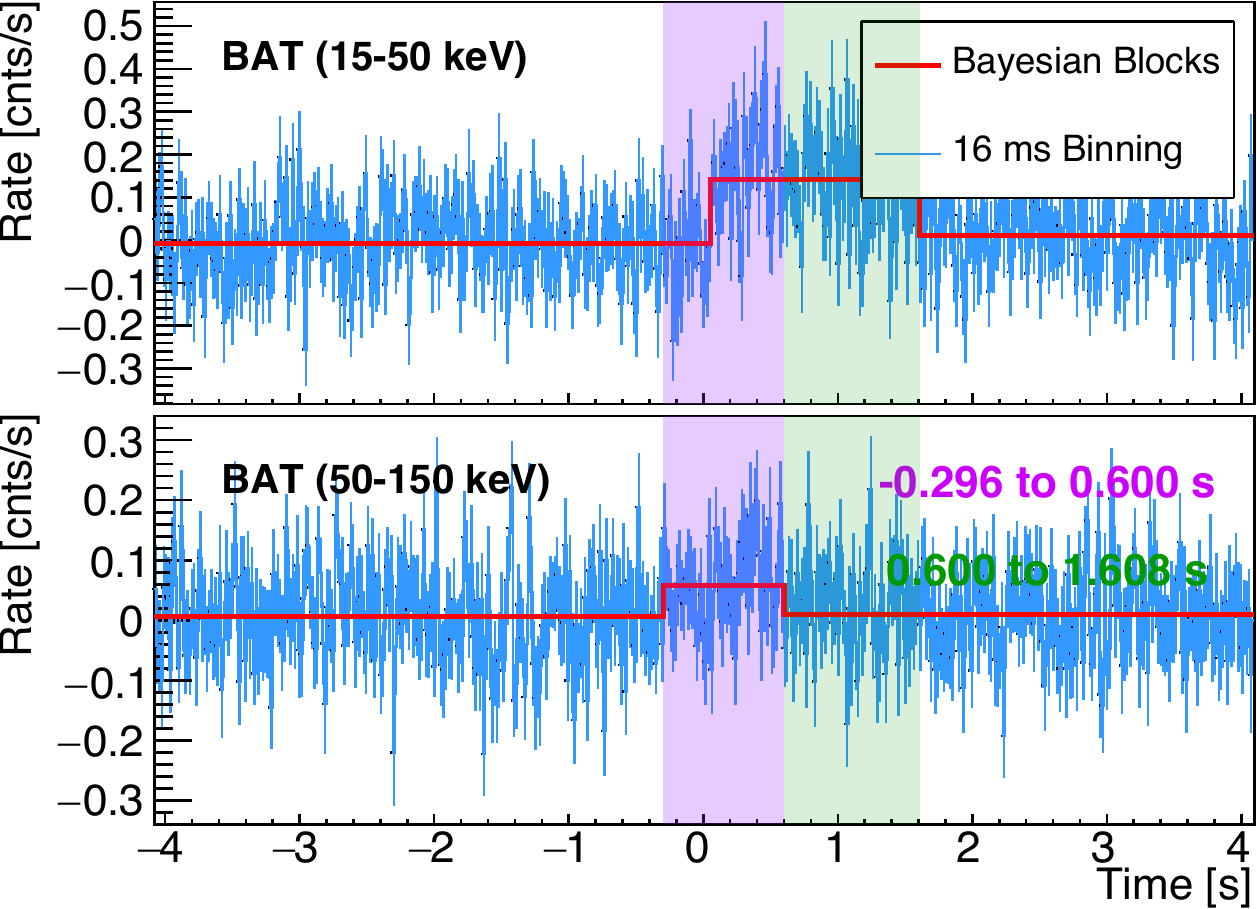}
      \label{fig:lc_100724A}
   }
   \subfigure[GRB~101219A]{
      \includegraphics[width=0.31\textwidth]{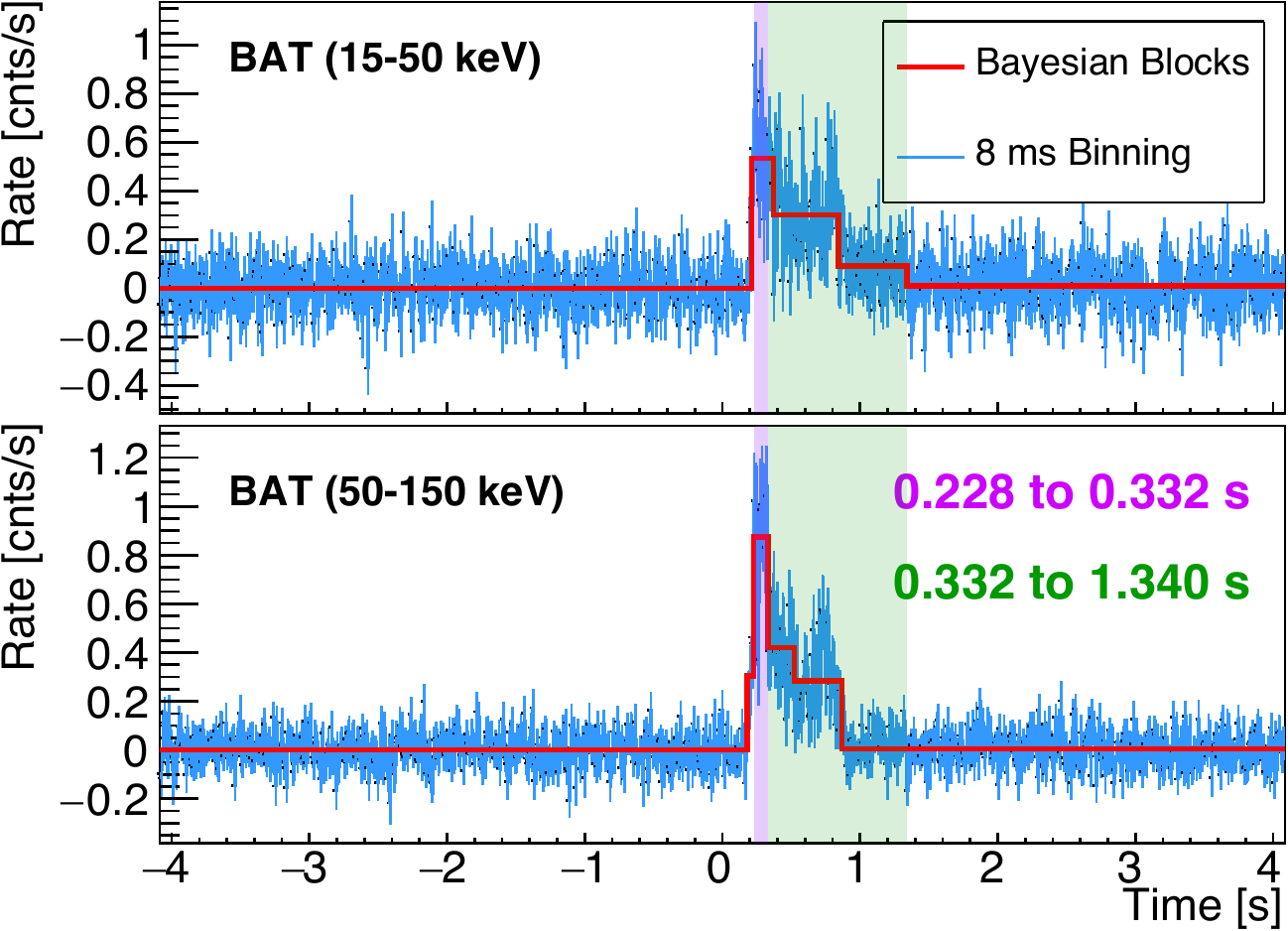}
      \label{fig:lc_101219A}
   }
   \subfigure[GRB~120804A]{
      \includegraphics[width=0.31\textwidth]{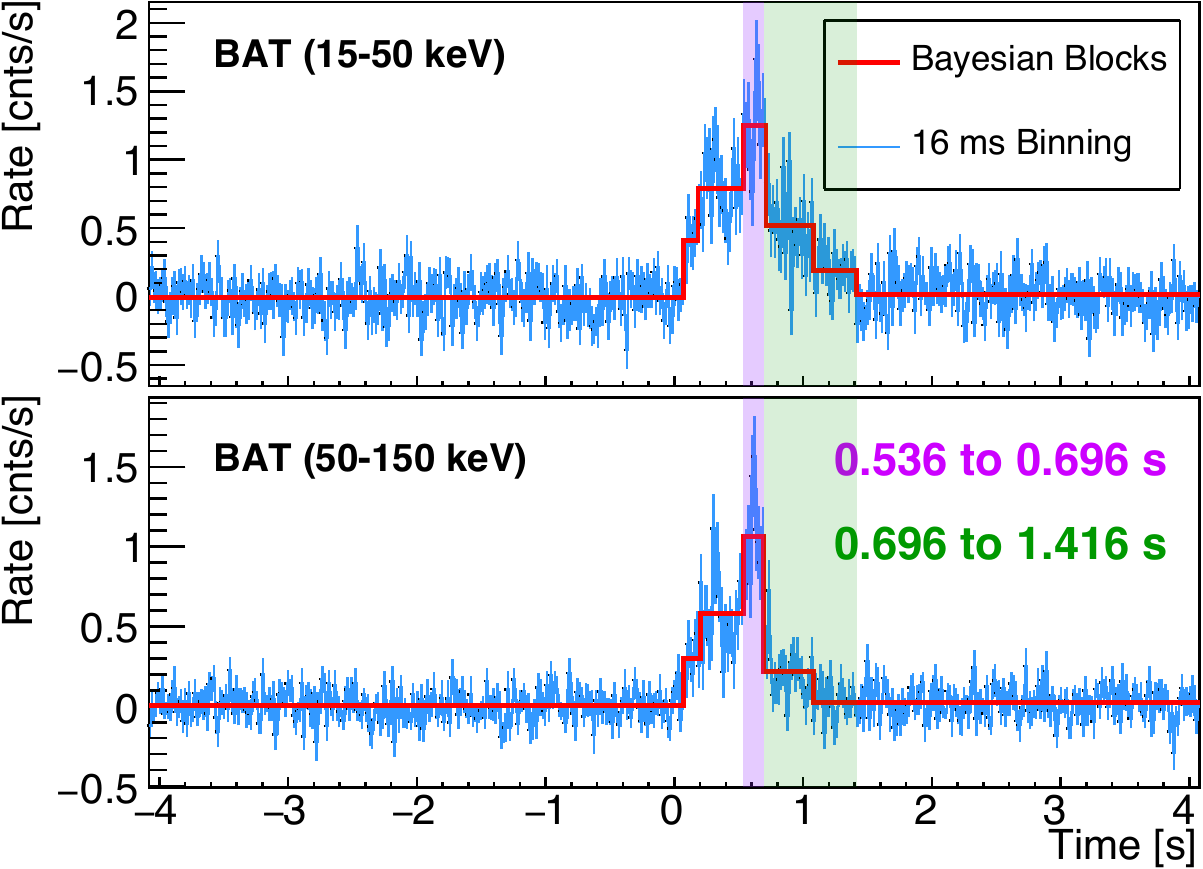}
      \label{fig:lc_120804a}
   }
   \subfigure[GRB~131004A]{
      \includegraphics[width=0.31\textwidth]{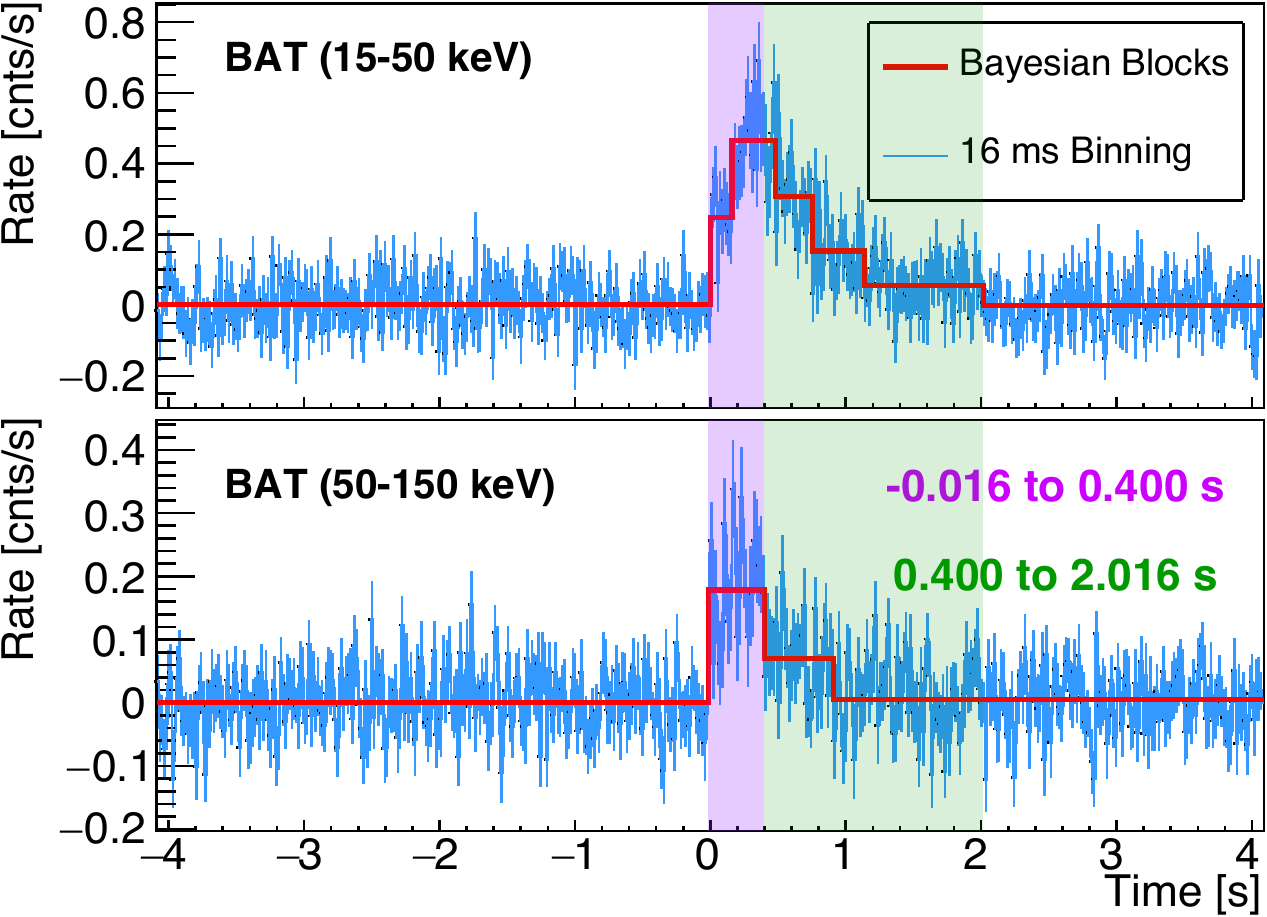}
      \label{fig:lc_131004a}
   }
   \subfigure[GRB~140903A]{
      \includegraphics[width=0.31\textwidth]{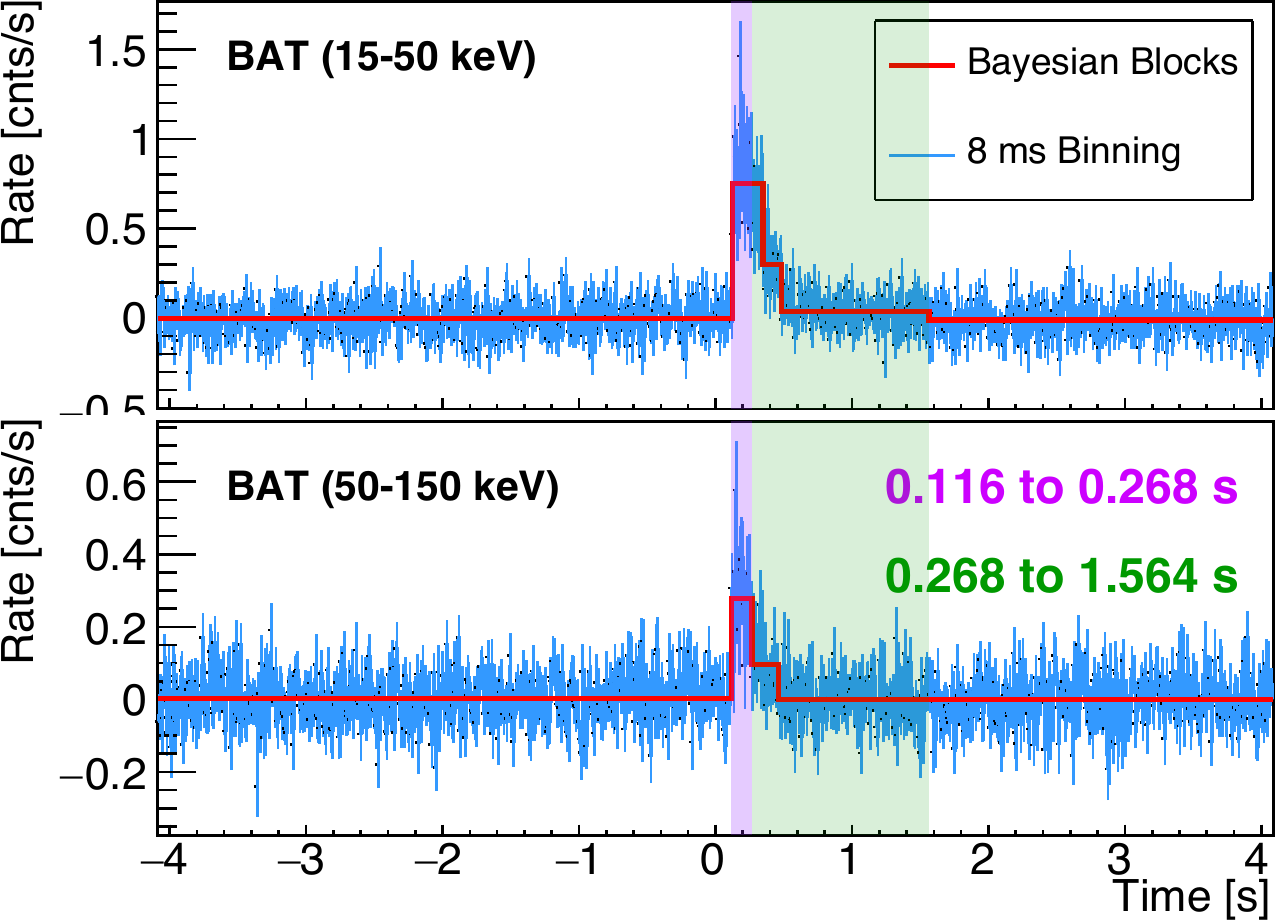}
      \label{fig:lc_140903a}
   }
   \subfigure[GRB~160821B]{
      \includegraphics[width=0.31\textwidth]{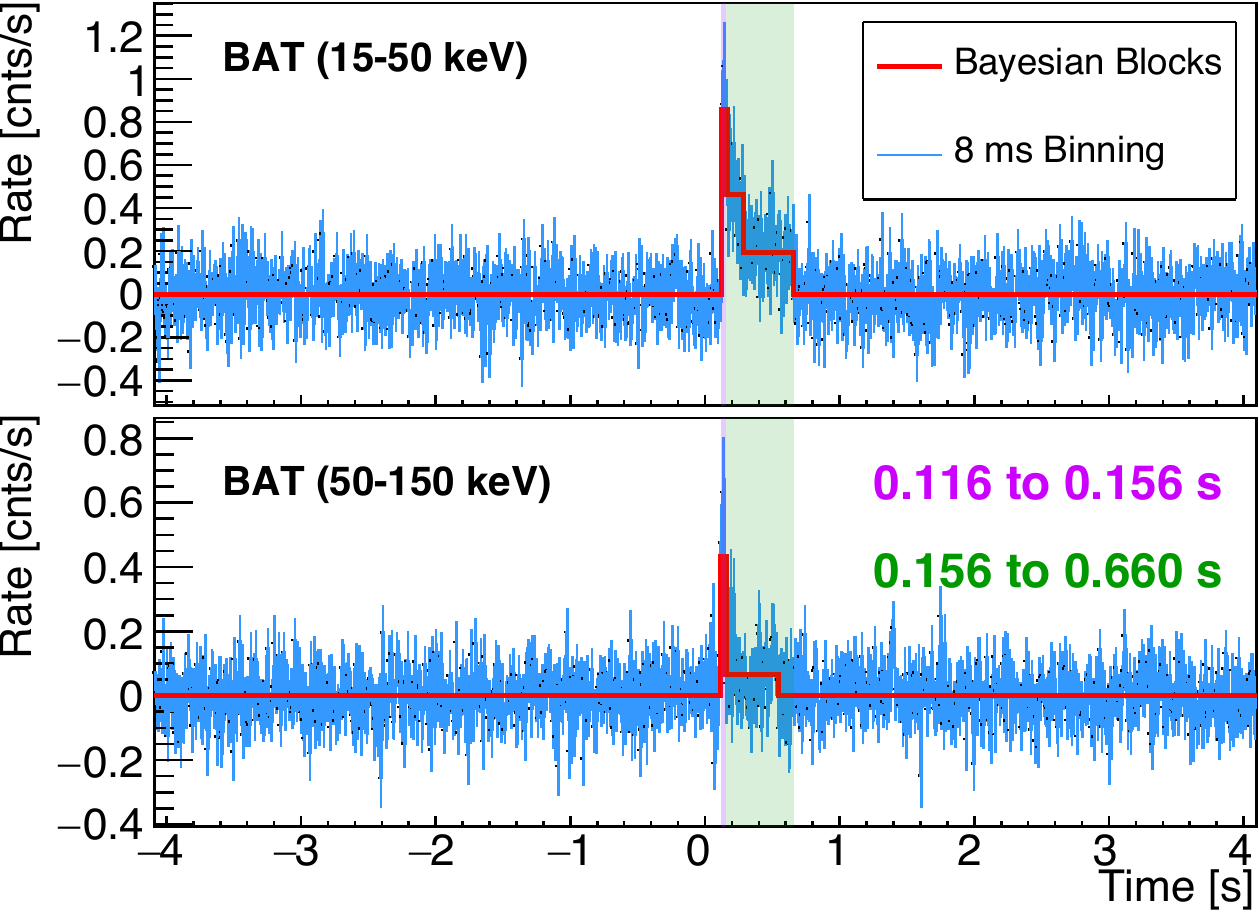}
      \label{fig:lc_160821b}
   }
   \subfigure[GRB~201221D]{
      \includegraphics[width=0.31\textwidth]{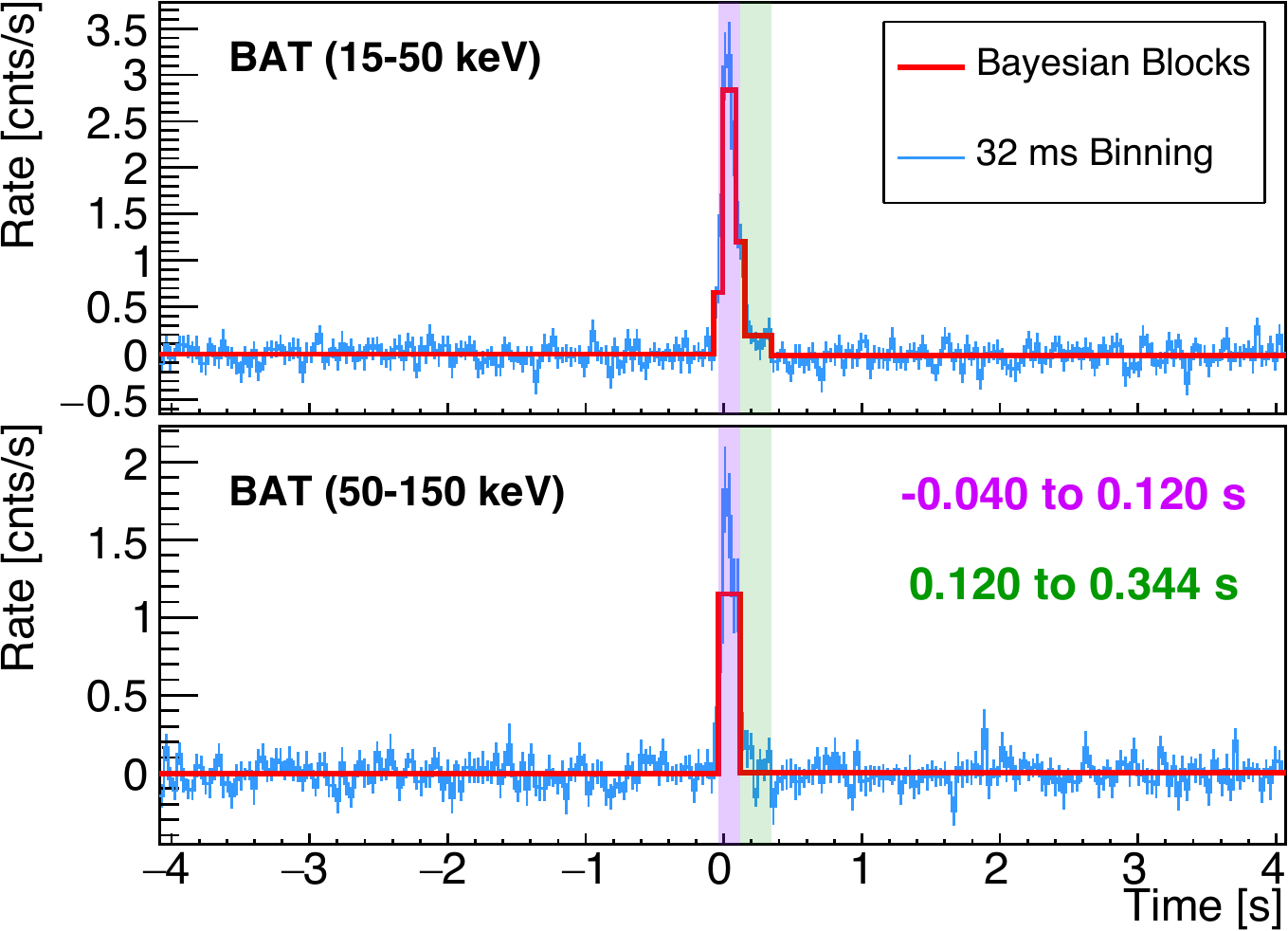}
      \label{fig:lc_201221d}
   }
   \subfigure[GRB~211023B]{
      \includegraphics[width=0.31\textwidth]{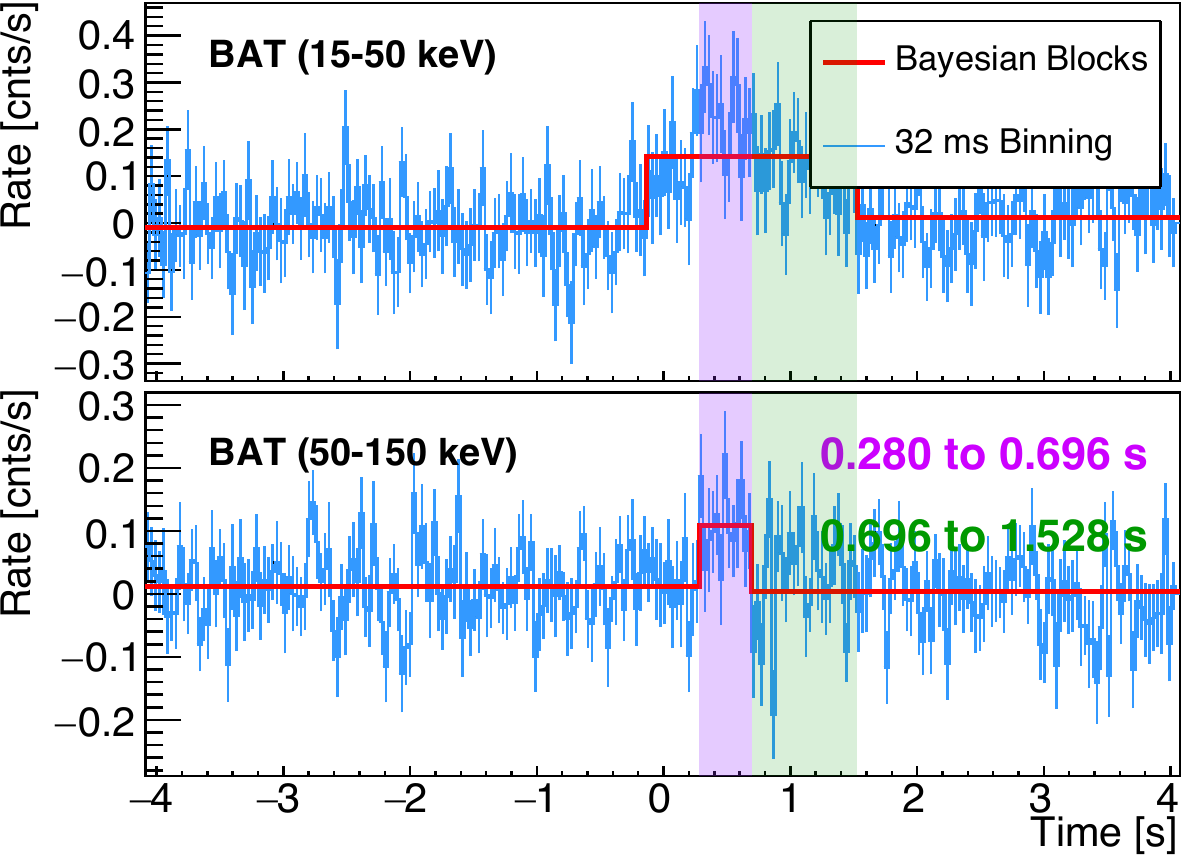}
      \label{fig:lc_211023B}
   }
   \caption{
      15--50\,keV (top panel) and 50--150\,keV (bottom panel) light curves detected by the {\swift}-BAT detector.
      The red lines represent the non-uniform bin light curves as determined by the bayesian block analysis.
      The brightest bin in the 50--150\,keV count rate is defined as the hard spike (purple shaded region),
      and the subsequent time region is defined as the soft tail (green shaded region), for which spectral analysis was conducted.
   }
   \label{fig:lc_all}
 \end{figure*}

In this sample of 43 events, events with multiple pulses\footnote{GRB~051221A, GRB~060313, GRB~080905A, GRB~100625A, GRB~160726A, and GRB~180204A}, such as GRB~060313 \citep{Roming+2006}, were removed manually as the soft tail cannot be identified.
Finally, a sample of 37 events remained.
Out of these 37 events, 9 have redshift measurements, and they were selected for spectral analysis in the next section.
Their light curves are shown in Figure~\ref{fig:lc_all}.
The top panel of Figure~\ref{fig:lc_all} shows the 15--50\,keV light curves, and the bottom panel shows the 50--150\,keV light curves.
The purple and green regions indicate the hard spike and soft tail, respectively, and their time intervals are shown in the bottom panel.
It should be noted that the signal to noise ratio is inconveniently low for some events, which is expected to affect the robustness of our results.

\subsection{Spectral Analysis}
Initially, we followed the ``Third {\swift} Burst Alert Telescope Gamma-Ray Burst Catalog'' \citep{Lien+2016} and fitted the data with a simple power law (PL) and a cutoff power law (CPL) models.
The PL model is represented by the following equation:
\begin{equation}
   N(E) = K_{\mathrm{PL}} \left(\frac{E}{50\,\mathrm{keV}}\right)^{\Gamma_\mathrm{ph}} .
\end{equation}
Here, $N(E)$ is the photon flux density at the energy $E$, $\Gamma_\mathrm{ph}$ is the photon index, and $K_\mathrm{PL}$ is the normalization factor at 50\,keV in units of photons~\si{cm^{-2}.s^{-1}.keV^{-1}}.
The CPL model is expressed by the equation:
\begin{equation}
   N(E) = K_{\mathrm{CPL}} \left(\frac{E}{50\,\mathrm{keV}}\right)^{\Gamma_\mathrm{ph}} 
          \exp \left[- \frac{\left(2+\Gamma_\mathrm{ph}\right)E}{\Epeak}\right] .
\end{equation} 
Here, $K_{\mathrm{CPL}}$ is the normalization factor at 50\,keV in units of photons~\si{cm^{-2}.s^{-1}.keV^{-1}}, and {\Epeak} is the peak energy in the ${\nu}F_{\nu}$ [i.e., $E^2N(E)$] spectrum, 
where $F_{\nu}=EN(E)$ represents the energy flux density.

According to \citet{Sakamoto+2011}, most of the GRB spectra observed by {\swift}/BAT can be well-fitted with the PL model, and changing to CPL does not significantly improve the fit.
Hence, we adopted the same criteria as \citet{Sakamoto+2011} and determined that CPL is more reasonable if $\Delta \chi^2 \equiv \chi_{\mathrm{PL}}^2 - \chi_{\mathrm{CPL}}^2 > 6$, where $\chi_{\mathrm{PL}}^2$ and $\chi_{\mathrm{CPL}}^2$ are the chi-square values for the PL and CPL model fits, respectively.

It is known that the soft tail of GRB~170817A can also be well-fitted with a blackbody (BB) model \citep{Goldstein+2017}.
Therefore, for the soft tail of the events selected in section \ref{subsec:lc}, we also performed fitting with the BB model, represented by the following equation:
\begin{equation}
   N(E) = \frac{K_{\mathrm{BB}}{\times}8.0525E^2\mathrm{d}E}{(k_{\mathrm{B}}T)^4\left[\exp\left(\frac{E}{k_{\mathrm{B}}T}\right)-1\right]} .
\end{equation}
Here, $k_{\mathrm{B}}T$ is the temperature in units of keV, and $K_{\mathrm{BB}}$ is the normalization in units of photons~\si{cm^{-2}.s^{-1}.keV^{-1}}.

For the spectral analysis, we used the widely adopted X-ray fitting tool  \texttt{XSPEC} version 12.12.0 \citep{Arnaud+1996}.
In addition, spectra were generated by dividing the 15--150\,keV range into 10 equally spaced energy bins on a logarithmic scale, followed by fitting.

Table~\ref{tab:spec} shows the results of the spectral analysis\footnote{We also identified a soft tail in GRB~050708 observed by {\hete} (Ogino, in preparation), which was added to our analysis.}.
Most of the soft tail spectra can be fitted with either one of the of the two non-thermal models (PL or CPL), or with the BB model.
However, there are 5 events, e.g., GRB~101219A, that could not be fitted well with the BB model. 
This may suggest that both the hard spike and the soft tail are non-thermal emissions from the same origin.
For events that are well-fitted with BB, the temperature is found as ${\sim}10$\,keV at the observer frame, yielding results consistent with \citet{Kienlin+2019}.

Furthermore, the energy flux $F$ was calculated for the 10--1000\,keV energy range, assuming that the spectral parameters defined at 15--150 keV extends up to 1000\,keV, and the isotropic luminosity \Liso was calculated using the formula $\Liso=4{\pi}d_\mathrm{L}^2F$.
Here, $d_\mathrm{L}$ represents the luminosity distance\footnote{Throughout this paper, we used $\Omega_m=0.286$, $\Omega_\Lambda=0.714$, $H_0=69.6$\,\si{km.s^{-1}.Mpc^{-1}} \citep{Bennett+2014}.}.

\begin{table*}[b]
   \tbl{Spectral analysis results for GRB~170817A-like events with known redshift}{
      \centering
      \scalebox{0.75}{
      \begin{tabular}{cccccccccc}
      \toprule
      GRB     & Redshift              & Time Int        & Model & ${\Gamma}_{ph}$         & {\Epeak}                 & $k_{\mathrm{B}}T$        & stat$/$dof            & Energy flux                          & \Liso                           \\ 
              &                       & [s]             &       &                         & [keV]                    & [keV]                    &                       & [\si{erg.s^{-1}.cm^{-2}}]            & [\si{erg.s^{-1}}]               \\ 
      \midrule
      060801  & 1.1304                & \ \ 0.038:0.162 & PL    & \Val{ 0.40}{0.37}{0.32} &                          &                          & $11.12/8$             & \VAL{5.01}{1.31}{1.42}{10^{-6}}      & \VAL{3.65}{0.96}{1.03}{10^{52}} \\
              & \citep{Berger+2007}   & \ \ 0.162:0.582 & PL    & \Val{ 0.41}{0.32}{0.29} &                          &                          & $ 4.33/8$             & \VAL{2.31}{0.43}{0.67}{10^{-6}}      & \VAL{1.68}{0.32}{0.49}{10^{52}} \\
              &                       & \ \ 0.162:0.582 & BB    &                         &                          & \Val{31.45}{6.92}{11.56} & $ 3.70/8$             &                                      &                                 \\
      100724A & 1.288                 &  $-$0.296:0.600 & PL    & \Val{ 1.26}{0.29}{0.28} &                          &                          & $ 5.94/8$             & \VAL{9.73}{5.67}{0.76}{10^{-7}}      & \VAL{9.81}{5.72}{0.77}{10^{51}} \\
              & \citep{Thoene+2010}   & \ \ 0.600:1.608 & PL    & \Val{ 2.05}{0.30}{0.33} &                          &                          & $11.65/8$             & \VAL{9.84}{5.58}{0.36}{10^{-8}}      & \VAL{9.92}{0.56}{3.68}{10^{50}}\\
              &                       & \ \ 0.600:1.608 & BB    &                         &                          & \Val{ 8.17}{1.31}{1.57}  & $ 7.03/8$             &                                      & \\
      101219A & 0.718                 & \ \ 0.228:0.332 & PL    & \Val{ 0.27}{0.16}{0.25} &                          &                          & $ 8.33/8$             & \VAL{3.60}{0.45}{0.46}{10^{-6}}      & \VAL{8.53}{1.05}{1.09}{10^{52}}\\
              & \citep{Fong+2013}     & \ \ 0.332:1.340 & PL    & \Val{ 0.81}{0.11}{0.11} &                          &                          & $ 7.37/8$             & \VAL{3.07}{0.24}{0.29}{10^{-7}}      & \VAL{7.27}{0.56}{0.70}{10^{51}}\\
              &                       & \ \ 0.332:1.340 & BB    &                         &                          & \Val{23.80}{1.34}{1.34}  & $32.00/8$             &                                      & \\
      120804A & 1.3                   & \ \ 0.536:0.696 & CPL   & \Val{ 0.26}{0.29}{0.58} & \Val{114.3}{28.4}{126.0} &                          & $ 2.88/7$             & \VAL{2.16}{0.14}{0.66}{10^{-6}}      & \VAL{2.23}{0.15}{0.68}{10^{52}}\\
              & \citep{Berger+2013}   & \ \ 0.696:1.416 & CPL   & \Val{ 0.54}{0.15}{0.15} & \Val{ 50.5}{ 6.1}{  6.1} &                          & $15.61/7$             & \VAL{3.12}{0.12}{0.12}{10^{-7}}      & \VAL{3.22}{0.12}{0.12}{10^{51}}\\
              &                       & \ \ 0.696:1.416 & BB    &                         &                          & \Val{11.84}{0.87}{0.87}  & $24.53/8$             &                                      & \\
      131004A & 0.717                 &  $-$0.016:0.400 & CPL   & \Val{-0.32}{0.79}{0.52} & \Val{ 71.0}{ 9.9}{ 12.9} &                          & $11.86/7$             & \VAL{3.16}{0.09}{0.30}{10^{-7}}      & \VAL{7.46}{0.22}{0.71}{10^{50}} \\
              & \citep{Chornock+2013} & \ \ 0.400:2.016 & CPL   & \Val{ 0.00}{0.28}{0.28} & \Val{ 36.3}{ 3.9}{  3.9} &                          & $18.76/7$             & \VAL{7.71}{0.27}{0.27}{10^{-8}}      & \VAL{1.82}{0.06}{0.06}{10^{50}} \\
              &                       & \ \ 0.400:2.016 & BB    &                         &                          & \Val{ 9.15}{0.55}{0.55}  & $21.71/8$             &                                      & \\
      140903A & 0.351                 & \ \ 0.116:0.268 & CPL   & \Val{ 0.56}{0.61}{0.65} & \Val{ 65.0}{11.3}{  7.5} &                          & $12.20/7$             & \VAL{6.35}{0.28}{0.37}{10^{-8}}      & \VAL{2.69}{0.12}{0.16}{10^{50}} \\
              & \citep{Troja+2016}    & \ \ 0.268:1.564 & CPL   & \Val{-0.66}{0.39}{1.11} & \Val{ 37.9}{ 5.1}{  6.2} &                          & $ 7.14/7$             & \VAL{5.34}{0.35}{0.34}{10^{-8}}      & \VAL{2.26}{0.15}{0.14}{10^{49}} \\
              &                       & \ \ 0.268:1.564 & BB    &                         &                          & \Val{ 9.52}{1.14}{1.29}  & $ 7.93/7$             &                                      & \\
      160821B & 0.162                 & \ \ 0.116:0.156 & CPL   & \Val{-0.41}{0.40}{0.40} & \Val{101.5}{21.7}{ 21.7} &                          & $14.74/7$             & \VAL{8.08}{0.55}{0.55}{10^{-7}}      & \VAL{5.91}{0.40}{0.40}{10^{49}} \\
              & \citep{Levan+2016}    & \ \ 0.156:0.660 & CPL   & \Val{-0.79}{0.32}{0.86} & \Val{ 40.5}{ 4.2}{  2.8} &                          & $12.06/7$             & \VAL{1.49}{0.08}{0.07}{10^{-7}}      & \VAL{1.09}{0.06}{0.05}{10^{49}} \\
              &                       & \ \ 0.156:0.660 & BB    &                         &                          & \Val{10.31}{0.97}{1.05}  & $12.05/7$             &                                      & \\
      201221D & 1.046                 &  $-$0.040:0.120 & CPL   & \Val{ 0.28}{0.24}{0.24} & \Val{ 84.0}{11.5}{ 44.6} &                          & $ 5.51/7$             & \VAL{2.67}{0.20}{0.34}{10^{-6}}      & \VAL{1.60}{0.12}{0.21}{10^{52}} \\
              & \citep{Postigo+2020}  & \ \ 0.120:0.344 & CPL   & \Val{ 2.07}{0.03}{0.03} & \Val{ 35.3}{11.9}{ 11.9} &                          & $ 9.68/7$             & \VAL{5.75}{1.90}{0.69}{10^{-7}}      & \VAL{3.45}{1.14}{0.42}{10^{51}} \\
              &                       & \ \ 0.120:0.344 & BB    &                         &                          & \Val{ 8.46}{2.17}{3.67}  & $15.56/7$             &                                      & \\
      211023B & 0.862                 & \ \ 0.280:0.696 & PL    & \Val{ 1.50}{0.25}{0.25} &                          &                          & $14.99/8$             & \VAL{7.16}{1.16}{1.19}{10^{-7}}      & \VAL{2.66}{0.43}{0.44}{10^{51}} \\
              & \citep{Rossi+2021}    & \ \ 0.696:1.528 & PL    & \Val{ 2.11}{0.39}{0.45} &                          &                          & $13.53/8$             & \VAL{1.50}{0.26}{0.29}{10^{-7}}      & \VAL{5.57}{0.97}{1.07}{10^{50}} \\
              &                       & \ \ 0.696:1.528 & BB    &                         &                          & \Val{ 7.54}{1.47}{1.93}  & $11.57/8$             &                                      & \\
      \midrule
      050709 & 0.16                   & \ \ 0.008:0.048 & CPL   & \Val{ 0.78}{0.34}{0.34} & \Val{210.0}{44.6}{115.5} &                          & $164/142$             & \VAL{5.67}{0.51}{0.54}{10^{-6}}      & \VAL{2.81}{0.09}{0.10}{10^{49}} \\
              & \citep{Hjorth+2005}   & \ \ 0.048:0.456 & CPL   & \Val{ 1.01}{0.55}{0.23} & \Val{120.3}{56.1}{ 57.3} &                          & $95.6/97$             & \VAL{7.25}{1.90}{1.69}{10^{-7}}      & \VAL{5.15}{0.50}{0.44}{10^{48}} \\
              &                       & \ \ 0.048:0.456 & BB    &                         &                          & \Val{34.6}{11.1}{13.6}   & $150.5/97$            &                                      & \\
      \midrule
      101224A & 0.4536                &  $-$0.256:0.256 & CPL   & \Val{-1.04}{0.39}{0.39} & \Val{341}{320}{320}      &                          &  $487.0/486$          & \VAL{4.04}{0.06}{0.06}{10^{-6}}      & \VAL{3.14}{0.04}{0.04}{10^{50}} \\
              & \citep{Fong+2022}     & \ \ 1.280:2.048 & CPL   & \Val{ 1.28}{0.86}{0.86} & \Val{37.2}{28.0}{16.0}   &                          &  $192.6/243$          & \VAL{1.25}{0.42}{0.39}{10^{-7}}      & \VAL{9.72}{0.32}{0.30}{10^{49}} \\
              &                       & \ \ 1.280:2.048 & BB    &                         &                          & \Val{7.63}{2.52}{2.52}   &  $194.0/244$          &                                      & \\
      150101B & 0.134                 &  $-$0.016:0.000 & CPL   & \Val{-0.80}{0.20}{0.20} & \Val{524}{176}{176}      &                          &  $638.2/885$          & \VAL{7.06}{0.31}{0.31}{10^{-6}}      & \VAL{3.41}{0.15}{0.15}{10^{50}} \\
              & \citep{Levan+2015}    & \ \ 0.000:0.064 & CPL   & \Val{-2.19}{0.51}{0.42} & \Val{21.9}{7.3}{7.3}     &                          &  $113.2/245$          & \VAL{3.34}{0.65}{0.32}{10^{-7}}      & \VAL{1.61}{0.31}{0.15}{10^{49}} \\
              &                       & \ \ 0.000:0.064 & BB    &                         &                          & \Val{5.62}{1.70}{3.24}   &  $114.4/246$          &                                      & \\
      170817A & 0.009783              &  $-$0.512:0.512 & CPL   & \Val{-0.84}{0.39}{0.39} & \Val{197}{89}{89}        &                          &  $527.3/506$          & \VAL{2.11}{0.21}{0.21}{10^{-7}}      & \VAL{4.55}{0.45}{0.45}{10^{46}} \\
              & \citep{Levan+2017}    & \ \ 0.512:2.048 & CPL   & \Val{-1.33}{0.48}{1.76} & \Val{40.0}{12.5}{14.8}   &                          &  $242.2/370$          & \VAL{3.99}{0.74}{0.61}{10^{-8}}      & \VAL{8.61}{1.60}{1.33}{10^{45}} \\
              &                       & \ \ 0.512:2.048 & BB    &                         &                          & \Val{10.3}{2.88}{3.52}   &  $243.8/371$          &                                      & \\
      \bottomrule
      \end{tabular}
      }
   }
   \label{tab:spec}
\end{table*}

\section{Correlation} \label{sec:correlation}
In this section, we assume the non-thermal model for all hard spikes and soft tails and we do not discuss the BB model.

There is a strong correlation between the rest-frame spectral peak energy {\Epeak} and the luminosity of GRBs, known as the Yonetoku relation \citep{Yonetoku+2004}.
Additionally, it is known that there is also a strong correlation in time-resolved spectra, name as the Golenetskii relation \citep{Golenetskii+1983}.
\citet{Lu+2012} calculated {\Liso} and {\Epeak} from time-resolved spectra using data from 14 {\LGRBs} and 1 {\SGRB} observed by {\fermi}, and found that the Golenetskii relation is consistent with the Yonetoku relation reported by \citet{Yonetoku+2010}.
Furthermore, the Golenetskii relation has been reproduced by some numerical simulations of the photospheric emission model (see also \cite{Lopez+2014,Parsotan+2018,Ito+2023}).

\begin{figure*}
   \centering
   \subfigure[Golenetskii relation]{
      \includegraphics[width=0.484\textwidth]{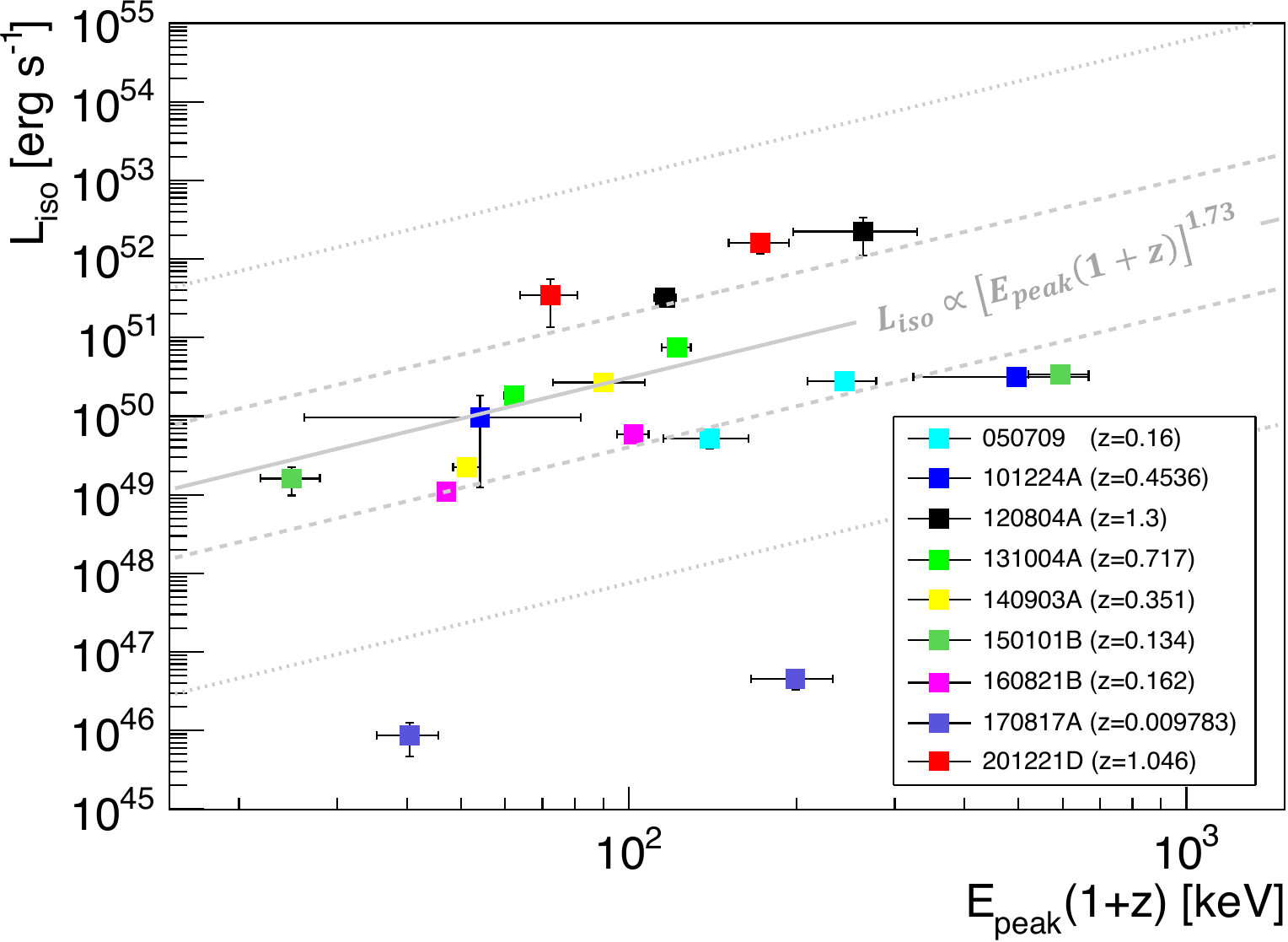}
      \label{fig:golenetskii}
   }
   \subfigure[Norm $A(z)$ of each GRBs]{
      \includegraphics[width=0.484\textwidth]{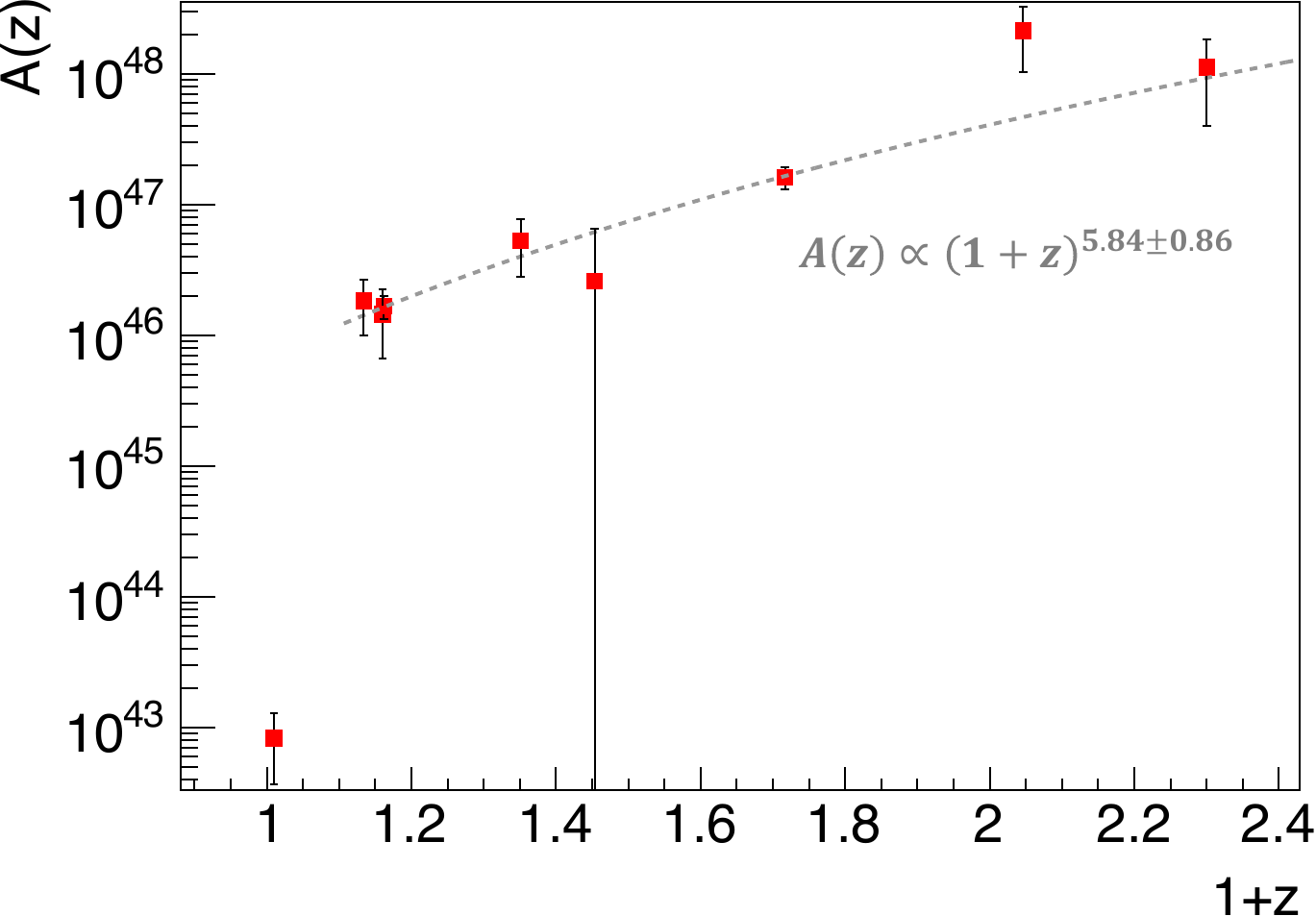}
      \label{fig:z_vs_norm}
   }
   \caption{
      (a) Golenetsky relation plotted for events with {\Epeak} of their hard spike and soft tail as shown in Table 1. 
      The solid line is the result of the Yonetoku relation taken from \citet{Zhang+2012}, the dashed and dotted lines indicate $1\sigma$ and $3\sigma$. 
      (b) The norm $A(z)$ of each GRB was obtained from (a) and correlated with the redshift. 
      The norm is found to show a dependence of $A(z)\propto(1+z)^{5.84\pm0.86}$.
   }
\end{figure*}

Figure~\ref{fig:golenetskii} shows events for which the peak energy {\Epeak} in the hard spike and soft tail has been determined in Table \ref{tab:spec}, on the {\Epeak}--{\Liso} plane.
The solid line in the figure represents the Yonetoku relation for {\SGRBs} as described in \citet{Zhang+2012}, with the dashed and dotted lines indicating the $1\sigma$ and $3\sigma$ distributions, respectively.
From this figure it is apparent that all SGRBs in our sample together with SGRBs in the sample of \citet{Kienlin+2019} (except GRB~170817A) are generally within the $1\sigma$ range.
In addition, this figure shows that all GRBs transition from the hard spike to the soft tail parallel to the solid line, suggesting that both the hard spike and soft tail emission have a same origin.
The emission mechanism of the hard spike and soft tails is not yet clear, but they could arise from the photosphere of the jet and the photosphere of the cocoon \citep{Kasliwal+2017,Lazzati+2017, Gottlieb+2018,Ioka+2019,Hamidani+2023b}.

Figure~\ref{fig:z_vs_norm} shows the results of fitting each GRB with $\Liso=A\left[{\Epeak}(1+z)\right]^{1.73}$.
Here, $A$ is a normalization as a free parameter, indicating the brightness of each GRB.
And the index of 1.73 is the value of the Yonetoku relation reported in \citet{Zhang+2012}.
This figure reveals a strong dependency of the value of $A$ on redshift.
Fitting the redshift dependency of $A$ with a power function $A(z)=A_{0}(1+z)^k$ yielded $A_0=(6.95\pm1.81)	\times 10^{45}$ and $k=5.84\pm0.86$ excluding GRB~170817A (the lowest data point in Figure \ref{fig:z_vs_norm}).

The increase in luminosity due to redshift is referred to as luminosity evolution $g_{k}(z)$ and is often expressed as a simple power function like $g_{k}(z)=(1+z)^{k}$ \citep{Petrosian+1992,Lloyd+2002}.
Various studies have reported values for $k$ in the case of {\SGRBs}, such as $k = 3.3_{-3.7}^{+1.7}$ \citep{Yonetoku+2014}, $k = 4.269\pm0.134$ \citep{Paul+2018}, $k = 4.47_{-0.29}^{+0.47}$ \citep{Zhang+2018}, and $k = 4.78_{-0.18}^{+0.17}$ \citep{Guo+2020}, generally considered to be around 3--5.
Our calculated value of $k=5.84\pm0.86$ is consistent with the high end of this range.
Therefore, in this paper, we recognize it as the luminosity evolution term $g_k(z)$ while there is a possibility that the trend on Fig. \ref{fig:z_vs_norm} is caused by the small number of sampling from large scattered data.

\section{Discussion} \label{sec:discussion}
We find that redshift dependence of time-resolved {\Epeak}--{\Liso} correlation.
This origin is considered to be ``intrinsically bright'' or ``energy concentrated with a narrower jet'' as the redshift increases.
Therefore, in this section, we use the maximum likelihood method to determine the jet opening angle, including the effect of redshift evolution.


\subsection{Calculation Method of Jet Opening Angle}
We simply assume a prompt-jet angular structure $f(\theta)$ so that the isotropic luminosity at $\theta=0^\circ$ is {\Lmax}.
The isotropic jet luminosity $L(\theta)$ at the angle of $\theta$ can be expressed as follows:
\begin{equation}
\label{eq:Liso}
L(\theta) = \Lmax \cdot f_0 f(\theta) .
\end{equation}
Here, $f_0$ is a normalization constant defined by the following equation:
\begin{equation}
\label{eq:f0}
f_0 = \frac{1}{\int_{0}^{\pi/2} f(\theta) \mathrm{d}\theta} .
\end{equation}
For an observed isotropic luminosity is {\Lobs}, we assume that the viewing angle {\thetaOBS} can be determined as
\begin{equation}
\thetaOBS = f^{-1} \left(\frac{\Lobs}{\Lmax f_0}\right) ,
\end{equation}
where $f^{-1}$ is the inverse function of $f$ defined in equation (\ref{eq:Liso}).
For instance, in the case of a Gaussian jet, the shape of the jet can be expressed as
\begin{equation}
f(\theta;\thetaC) = \exp\left(-\frac{\theta^2}{2\thetaC^2}\right) .
\end{equation}
Here, $\thetaC$ is the core angle of the jet, representing the opening angle of the Gaussian jet.
From the observed isotropic luminosity {\Lobs}, the viewing angle {\thetaOBS} is calculated as follows:
\begin{equation}
\label{eq:thetaOBS}
\thetaOBS = \sqrt{-2\thetaC^2 \ln \left(\frac{\Lobs}{\Lmax f_0}\right)} .
\end{equation}
In this paper, we assume the Gaussian jet and aim to derive its opening angle {\thetaC} using the maximum likelihood method.
The computational procedure is as follows.
\begin{enumerate}
\item We consider a Gaussian jet structure with the opening angle {\thetaC} and the maximum luminosity \Lmax.
\item From equation (\ref{eq:thetaOBS}), we determine the viewing angle {\thetaOBS} from the observed luminosity {\Lobs}.
\item We calculate the probability $p(\thetaC,\thetaOBS,\zobs)$ of observing the GRB with the opening angle {\thetaC} at redshift {\zobs}, at the angle {\thetaOBS}.
\item We compute $p(\thetaC,\thetaOBS,\zobs)$ for all events and calculate the likelihood $\displaystyle P(\thetaC)={\prod_i}p(\thetaC,\thetaOBS^{i},\zobs^{i})$.
\item We find the {\thetaC} that maximizes $P(\thetaC)$ which is the most plausible opening angle of the jet.
\end{enumerate}
Note that the {\Lmax} was set to be 10 times brighter than the one of GRB~120804A, the brightest in our sample (${\Lmax}=2.2{\times}10^{53}$\,\si{erg.s^{-1}}).
The viewing angle {\thetaOBS} used in the likelihood calculation is proportional to $[\ln(1/{\Lmax})]^{1/2}$, indicating that it is a function quite insensitive to the value of {\Lmax}. 
Therefore, the fixed value of {\Lmax} does not significantly affect the results.
The specific calculations are explained in the following section.

\subsection{Non-evolving Jet Opening Angle}
Previously it has been suggested that the opening angle of {\LGRBs} is anti-correlated with redshift \citep{Yonetoku+2005,Ronning+2019,Ronning+2020}.
The origin of this effect remains unclear, but it has been speculated to be due to the dependence on the metallicity and density of LGRB-progenitors on redshift.
However, as we are currently considering {\SGRBs} originating from the neutron star merger, 
it is natural to assume that their opening angle does not depend on redshift.

Thus, we first assume that the maximum luminosity and the opening angle are independent of redshift,
and constant as ${\Lmax}\equiv{\LmaxO}$ and ${\thetaC}\equiv{\thetaCO}$, respectively.
In this case, the shape of the jet can be expressed as follows:
\begin{equation}
\label{eq:non-evo-jet}
\frac{L(\theta)}{g_k(z)} = \LmaxO \cdot f_0 \exp\left(-\frac{\theta^2}{2\thetaCO^2}\right) .
\end{equation}
By dividing the luminosity $L(\theta)$ by the luminosity evolution $g_k(z)$, we cancel out the dependency on redshift.
Therefore, using equation (\ref{eq:thetaOBS}), the observed redshift {\zobs}, and the observed isotropic luminosity {\Lobs}, the viewing angle {\thetaOBS} can be expressed as:
\begin{equation}
\thetaOBS = \sqrt{-2\thetaCO^2 \ln \left(\frac{\Lobs/g_k(\zobs)}{\LmaxO f_0}\right)} .
\end{equation}

Next, we consider the probability $p$ of detecting {\SGRBs} with luminosity $L$ occurring at redshift $z$.
Using the number of {\SGRBs} occurring at redshift $z$, denoted as $N(z)$, and the number of {\SGRBs} observed at any viewing angle {\thetaOBS}, denoted as {\Nobs}, $p$ can be written as follows.
\begin{equation}
   p \propto \frac{\Nobs}{N(z)} 
   \label{eq:p}
\end{equation}
Here, $N(z)$ is given by the following equation.
\begin{equation}
   N(z) \propto \Psi(\Llim(z)/g_k(z)) \times \rho(z,\thetaC)\left(\frac{\mathrm{d}V}{\mathrm{d}z}\right)_{z} \frac{1}{1+z}
   \label{eq:N_z}
\end{equation}
Here, $\left(\frac{\mathrm{d}V}{\mathrm{d}z}\right)_{z}$ is the differential comoving volume at redshift $z$,
and the functions $\Psi(\Llim(z)/g_k(z))$ and $\rho(z,\thetaC)$ represent the cumulative luminosity function and the formation rate of {\SGRBs}, respectively.
$\Llim(z)$ is the minimum observable luminosity determined by the flux limit of the detector {\Flim} and redshift $z$,
calculated as $\Llim = 4{\pi}d_\mathrm{L}^2{\Flim}$.
In this study, {\Flim} was set at $3\times10^{-8}$\,\si{erg.cm^{-2}.s^{-1}} for {\swift}/BAT \citep{Lien+2016},
and $3\times10^{-7}$\,\si{erg.cm^{-2}.s^{-1}} for {\fermi}/GBM \citep{Veres+2019}.
\citet{Yonetoku+2014} and \citet{Zhang+2018}, among others, discussed the formation rate ${\rho}_{\rm iso}(z)$ of {\SGRB} under the assumption of isotropic emission.
In this paper, we assume that the realistic formation rate $\rho(z,\thetaC)$ is expressed by the following equation:
\begin{equation}
   \rho(z,\thetaC) = \rho_{\rm iso}(z) / F(\thetaC) .
   \label{eq:rho_z}
\end{equation}
Here, $F(\thetaC)$ is the geometrical correction factor of jet collimation.
 Assuming that the majority of energy are contained in ${\theta}<{\thetaC}$, it defined as
\begin{equation}
   F(\thetaC) = \frac{2\int_{0}^{\thetaC}\sin{\theta}d{\theta}\int_{0}^{2\pi}d\phi}{4\pi} 
              \simeq \frac{\thetaC^2}{2}  .
   \label{eq:F}
\end{equation}

And, the number {\Nobs} of {\SGRBs} observed at any viewing angle {\thetaOBS} is proportional to the solid angle $\Delta\Omega$, and can be expressed as follows:
\begin{equation}
   {\Nobs} {\propto} \frac{\Delta\Omega}{4\pi} 
           {\propto} \sin{\thetaOBS} \Delta{\thetaOBS} .
   \label{eq:Nobs}
\end{equation}
Using the equations (\ref{eq:p}), (\ref{eq:N_z}), (\ref{eq:rho_z}), and (\ref{eq:Nobs}), $p$ can be written as follows:
\begin{equation}
   p(\thetaC,\thetaOBS,\zobs) \propto \frac{F(\thetaC)\sin\thetaOBS\Delta\thetaOBS}{\Psi({\Llim(\zobs)/g_k(z)}) \times \rho_{\rm iso}(\zobs)\left(\frac{\mathrm{d}V}{\mathrm{d}z}\right)_{\zobs} \frac{1}{1+\zobs}} .
\end{equation}
Thus, the likelihood $P(\theta_{\rm c,0})$, assuming $\Delta \thetaOBS$ is constant, can be expressed as
\begin{equation}
   \label{eq:non-evo-P}
   P({\theta_{\rm c,0}}) \propto \prod_{i=1}^{N} \frac{F(\theta_{\rm c,0})\sin\thetaOBS^{i}}{\Psi({\Llim(\zobs^{i})/g_k(\zobs^{i})}) \times \rho_{\rm iso}(\zobs^{i})\left(\frac{\mathrm{d}V}{\mathrm{d}z}\right)_{\zobs^{i}} \frac{1}{1+\zobs^{i}}} .
\end{equation}
This calculation used the luminosity function and the formation rate are used as reported in \citet{Zhang+2018}.

The result of this calculation gives {\thetaCO} as $\Val{32.4^{\circ}}{0.63^{\circ}}{0.63^{\circ}}\,$.
The opening angle is also obtained from observation of the jet break at the GRB afterglow phase \citep{Rhoads1999,Sari+1999}.
According to \citet{Escorial+2023}, jet break measurements reveal that there is a population of {\SGRBs} with wide jet opening angles ($\gtrsim 10^\circ$); 
overall the jet opening angle takes values in the range $\sim 0.5^\circ - 26^\circ$, and the average value is $\sim6^{\circ}$.
Assuming that the jet opening angle at the prompt emission phase and the afterglow phase is roughly the same\footnote{It is not obvious that the opening angle at the prompt and afterglow phase is the same, but we simply assume that they are the same and discuss them in this paper.}, our estimation indicates larger values on average.
Therefore, in the following, to avoid such large values, we introduce a more complex model where the opening angle is assumed to evolve with redshift. 

\subsection{Evolving Jet Opening Angle}
\begin{figure*}
\centering
\subfigure[Maximum luminosity evolution]{
\includegraphics[width=0.484\textwidth]{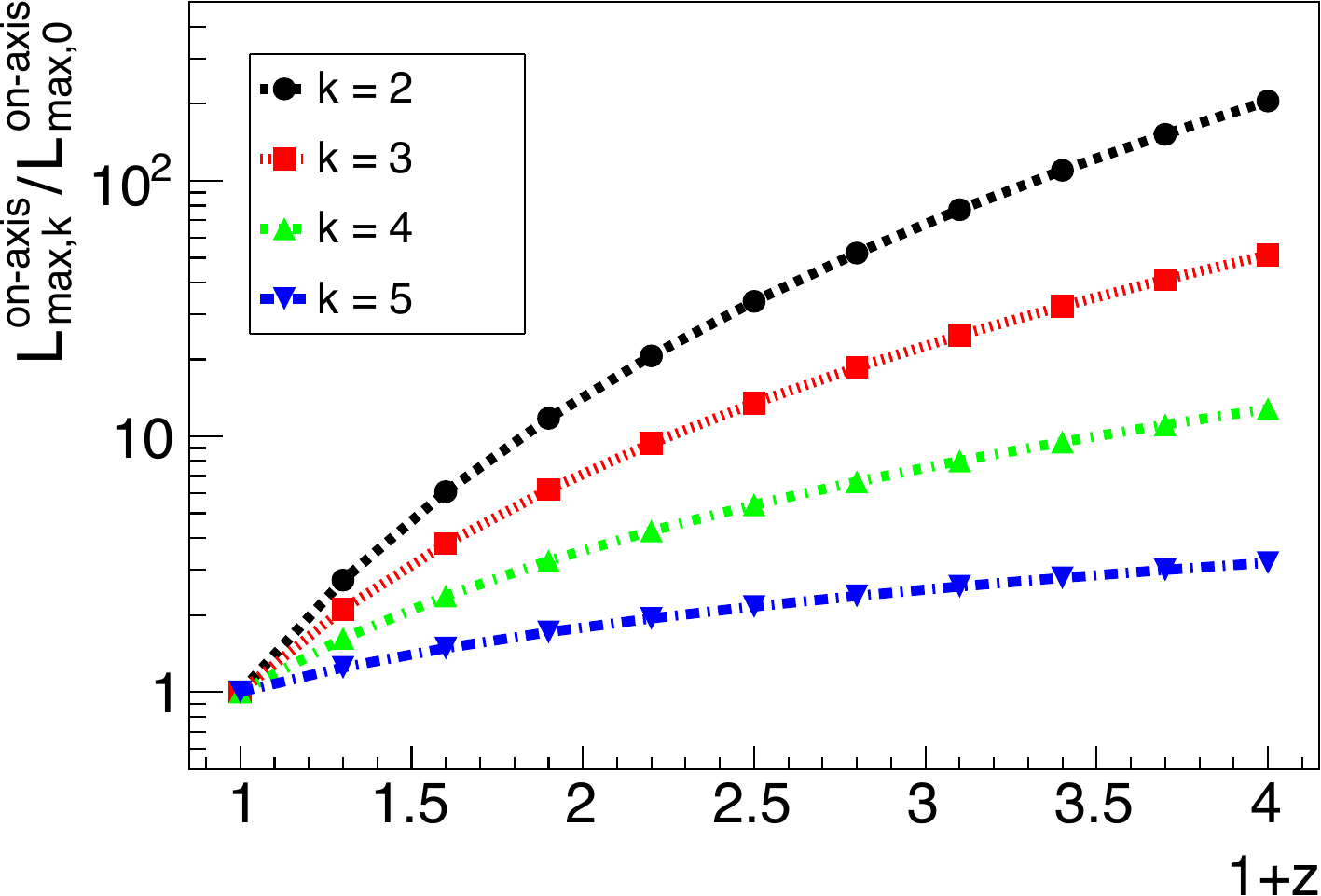}
\label{fig:max_lumi_evo}
}
\subfigure[Jet opening angle evolution]{
\includegraphics[width=0.484\textwidth]{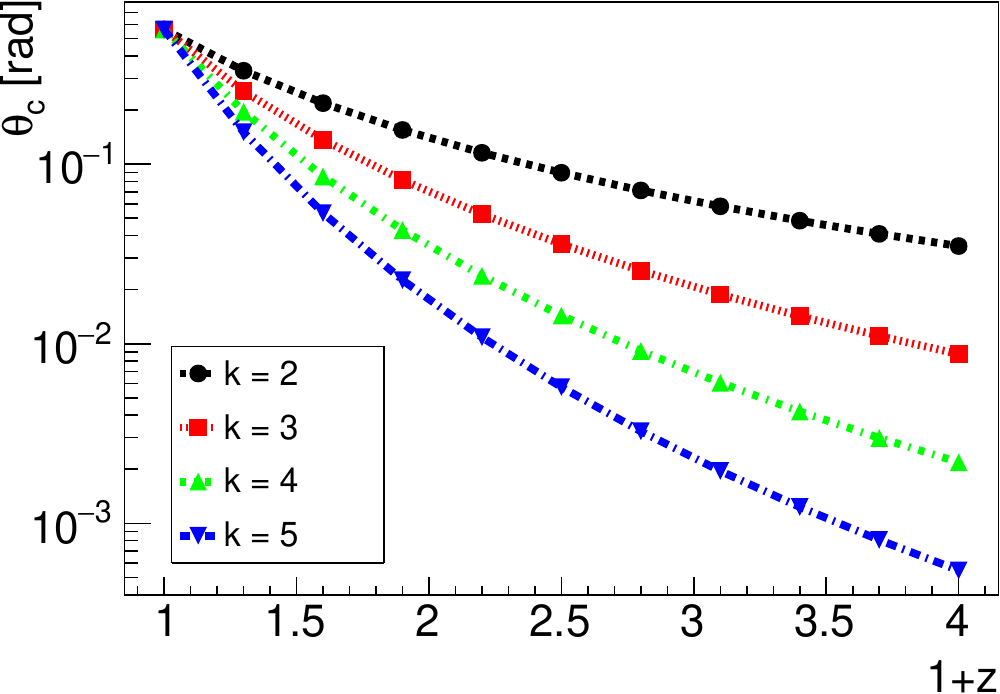}
\label{fig:theta_c_evo}
}
\caption{
   (a) Maximum luminosity evolution (normalized to maximum luminosity at $z=0$) and (b) jet opening angle evolution as a function of redshift $z$ (see equations (\ref{eq:max_lumi_evo}) and (\ref   {eq:f_evo})). Note that $k$ is the parameter expressing the strength of the redshift dependence.
}
\end{figure*}

We assume that the maximum luminosity {\Lmax} and the opening angle {\thetaC} depends on redshift,
and are denoted as ${\Lmaxk}(z)$ and ${\thetaCk}(z)$, respectively.
The normalization constant is also redshift dependent, denoted as {\fOk}.
In this case, the shape of the jet can be expressed as follows:
\begin{equation}
\label{eq:evo-jet}
L(\theta) = \Lmaxk(z) \cdot \fOk(z) \exp\left(-\frac{\theta^2}{2\thetaCk^2(z)}\right) .
\end{equation}
Using equations (\ref{eq:non-evo-jet}) and (\ref{eq:evo-jet}), solving for the luminosity evolution $g_k(z)$ yields
\begin{equation}
g_k(z) = \frac{\Lmaxk(z)}{\LmaxO} \cdot
\frac{\fOk(z)\exp\left(-\frac{\theta^2}{\thetaCk^2(z)}\right)}{f_0\exp\left(-\frac{\theta^2}{\thetaCO^2}\right)} .
\end{equation}
By setting $\theta=0^{\circ}$, the following transformation can be made:
\begin{equation}
g_k(z) = \frac{\Lmaxk^{\mathrm{on\mbox{-}axis}}(z)}{\LmaxO^{\mathrm{on\mbox{-}axis}}} \cdot \frac{\fOk(\thetaCk(z))}{f_0(\thetaCO)} = (1+z)^{5.84} .
\end{equation}
To emphasize that the values are at ${\theta}=0^{\circ}$, ``on-axis'' is written in superscript.
We separate the evolution of the maximum luminosity and the jet shape as follows:
\begin{equation} \label{eq:max_lumi_evo}
\frac{\Lmaxk^{\mathrm{on\mbox{-}axis}}(z)}{\LmaxO^{\mathrm{on\mbox{-}axis}}} = (1+z)^{5.84-k} ,
\end{equation}

\begin{equation} \label{eq:f_evo}
\frac{\fOk(\thetaCk(z))}{f_0(\thetaCO)} = (1+z)^{k} .
\end{equation}
As $\LmaxO^{\mathrm{on\mbox{-}axis}}$ and $f_0(\thetaCO)$ have already been determined in the previous section,
the numerical calculations can be performed to compute the evolution of the maximum luminosity and the opening angle.
The results are shown in Figure \ref{fig:max_lumi_evo} and Figure \ref{fig:theta_c_evo}.
From these results, it can be seen that in the range of $k=$3--5, at the typical redshift of \SGRBs, $z=0.72$ \citep{Kisaka+2017}, the average opening angle of $\sim6^{\circ}(0.1\,\mathrm{rad})$ is reasonably explained.

Hence, we assume that the opening angle $\thetaC(z)$ at any redshift can be expressed as the opening angle at $z=0$ weighted by {\thetaCO} and {\thetaCk}.
The equation is as follows:
\begin{equation}
\label{eq:thetaCZ}
\thetaC(z) = \thetaC(0) \times \left(\frac{\thetaCk(z)}{\thetaCO}\right) .
\end{equation}
Then, by modifying equation (\ref{eq:non-evo-P}), we determine the opening angle $\thetaC(0)$ using the maximum likelihood method, which allows us to calculate the opening angle at any redshift.

\begin{equation}
\label{eq:likelihood-z}
P[\thetaC(0)] \propto \prod_{i=1}^{N} \frac{F\left(\thetaC(0)\times\frac{\thetaCk(\zobs^{i})}{\thetaCO}\right)\sin\thetaOBS^{i}{\times}g_k(\zobs)}{\Psi({\Llim(\zobs^{i})/g_k(\zobs^{i})}) \times \rho_{\rm iso}(\zobs^{i})\left(\frac{\mathrm{d}V}{\mathrm{d}z}\right)_{\zobs^{i}} \frac{1}{1+\zobs^{i}}}
\end{equation}
\normalsize
Varying the value of $k$ from 3 to 5, the results of computing the opening angle {\thetaC} and viewing angle {\thetaOBS} using equations (\ref{eq:thetaCZ}) and (\ref{eq:likelihood-z}) are shown in Table \ref{tab:thetaC}.
Compared to the opening angle of $\sim6^{\circ}$ determined from the jet break \citep{Escorial+2023}, it is estimated that the evolution represented by equations (\ref{eq:max_lumi_evo}) and (\ref{eq:f_evo}) requires $k=\mbox{4--5}$.
Note that, for the case GRB~170817A, our model yields an unphysically large value of the viewing angle (as $>90^{\circ}$ compared to 20--30$^{\circ}$ \citep{Mooley+2018,Troja+2019}).
According to our model, the wide jet events of ${\thetaC}{\sim}30^{\circ}$ is expected at $z{\sim}0$, 
but GRB~170817A is found to be the narrow jet event with ${\thetaC}\sim3^{\circ}$ \citep{Troja+2019}. 
This suggests that GRB~170817A was detected as the very low-luminosity event which can not be explained by luminosity evolution.

\begin{table*}
   \tbl{Opening angle and viewing angle for each {\SGRBs}.}{
      \centering
      \scalebox{0.93}{
      \begin{tabular}{cccccccccc} 
         \toprule
                 &          &                                             & \multicolumn{2}{c}{$k=3$}                         & \multicolumn{2}{c}{$k=4$}                         & \multicolumn{2}{c}{$k=5$}                           & {\thetaC} \\
         \cmidrule[0.2pt](lr){4-9}                                                                                                                                                                                                    
         GRB     & Redshift & {\thetaCO} [$^{\circ}$]                     & {\thetaC} [$^{\circ}$]  & {\thetaOBS} [$^{\circ}$]& {\thetaC} [$^{\circ}$]  & {\thetaOBS} [$^{\circ}$]& {\thetaC} [$^{\circ}$]  & {\thetaOBS} [$^{\circ}$]  & from jet break [$^{\circ}$] \\ 
         \midrule                                                                                                                                                                                                     
         120804A & 1.3      & {\multirow{8}{4em}{\Val{32.4}{0.63}{0.63}}} & \Val{ 2.65}{0.04}{0.05} & \Val{ 8.49}{0.22}{0.22} & \Val{ 1.16}{0.02}{0.02} & \Val{ 3.12}{0.04}{0.05} & \Val{ 0.51}{0.01}{0.01} & \Val{ 1.85}{0.04}{0.05} & $>$\Val{10.5}{1.0}{1.2}     \\
         201221D & 1.046    &                                             & \Val{ 3.76}{0.06}{0.07} & \Val{12.02}{0.29}{0.29} & \Val{ 1.84}{0.04}{0.04} & \Val{ 4.89}{0.07}{0.08} & \Val{ 0.90}{0.02}{0.02} & \Val{ 3.25}{0.07}{0.07} & -                           \\
         131004A & 0.717    &                                             & \Val{ 6.36}{0.12}{0.13} & \Val{24.86}{0.36}{0.36} & \Val{ 3.70}{0.07}{0.07} & \Val{12.71}{0.24}{0.25} & \Val{ 2.16}{0.04}{0.04} & \Val{ 9.01}{0.13}{0.14} & -                           \\
         101224A & 0.4536   &                                             & \Val{10.47}{0.20}{0.21} & \Val{41.91}{0.57}{0.58} & \Val{ 7.21}{0.15}{0.16} & \Val{25.14}{0.46}{0.47} & \Val{ 4.95}{0.09}{0.10} & \Val{20.73}{0.32}{0.33} & -                           \\
         140903A & 0.351    &                                             & \Val{13.03}{0.24}{0.26} & \Val{52.00}{0.73}{0.74} & \Val{ 9.65}{0.19}{0.19} & \Val{33.31}{0.59}{0.60} & \Val{ 7.14}{0.13}{0.14} & \Val{29.55}{0.46}{0.46} & \Val{3.2}{2.0}{0.8}         \\
         160821B & 0.162    &                                             & \Val{20.47}{0.41}{0.41} & \Val{79.52}{1.13}{1.14} & \Val{17.62}{0.36}{0.37} & \Val{57.82}{1.04}{1.04} & \Val{15.16}{0.30}{0.30} & \Val{50.25}{0.10}{0.11} & \Val{8.4}{3.2}{4.9}         \\
         050709  & 0.16     &                                             & \Val{20.57}{0.40}{0.41} & \Val{86.98}{1.25}{1.26} & \Val{17.74}{0.37}{0.38} & \Val{65.33}{1.20}{1.21} & \Val{15.29}{0.29}{0.30} & \Val{55.48}{0.96}{0.97} & $>$\Val{25.8}{2.6}{2.2}     \\
         150101B & 0.134    &                                             & \Val{22.01}{0.43}{0.44} & \Val{83.61}{1.33}{1.35} & \Val{19.46}{0.39}{0.40} & \Val{61.60}{1.10}{1.11} & \Val{17.11}{0.35}{0.36} & \Val{56.14}{1.19}{1.19} & $>$\Val{9.4}{2.2}{2.8}      \\
         \bottomrule
      \end{tabular}
      }
   }
   \label{tab:thetaC}
\end{table*}

\section{Conclusion} \label{sec:conclusion}
Focusing of the soft tail in GRB~170817A, we used the {\swift}/BAT Burst Catalog to search for similar events with known redshift in {\SGRBs}. 
As a result, we discovered 9 new candidates. 
Next, we focused on events with identified {\Epeak} and examined the {\Epeak}--{\Liso} correlation. 
Then, we found that all events from the hard spike to the soft tail vary in time along the {\Epeak}--{\Liso} correlation.
This parallel time variation, and the fact that all spectra are well fitted by the non-thermal model, suggests that the hard spike and soft tail are emissions from the same origin.

We showed the redshift dependence of the time resolved {\Epeak}--{\Liso} correlation for selected events, which can be explained by the luminosity evolution of the {\SGRBs} as shown in previous studies.
Furthermore, we calculated the average jet opening angle for {\SGRBs} with redshift and {\Liso} measurements. 
We followed two models (independent or dependent on the redshift) to determine the jet structure and calculate the jet opening angle.
In the redshift-independent model, the jet angle was found as $\Val{32.4^\circ}{0.63^\circ}{0.63^\circ}\,$, which confirms the existence of {\SGRBs} with wide jets but is larger, on average, than jet opening angles derived from jet breaks as $\sim6^{\circ}$ \citep{Escorial+2023}. 
Using the redshift-dependent model, the consistency of our jet opening angles with jet opening angles derived from jet breaks could be improved for most events.
We also showed that GRB~170817A can be interpreted as the very faint peculiar event detected due to the observation of the narrow jet event at low redshift.
Our procedure here of estimating the jet structure can be expanded using other models for the jet structure (e.g., top-hat, power-law, etc.) and with more future {\SGRBs}, it is expected to improve our understanding of {\SGRBs} jets and their structure.

\begin{ack}
   This study was supported by JSPS KAKENHI Grant Numbers JP22J12717 (N.O.),
   JP17H06362 (M.A.), JP23H04898 (D.Y. and M.A.), JP23H04895 (T.S.), the CHOZEN Project of Kanazawa University (D.Y., M.A., and T.S.), 
   and the JSPS Leading Initiative for Excellent Young Researchers Program (M.A.).
\end{ack}

\bibliography{main}

\begin{thebibliography}{}
\expandafter\ifx\csname natexlab\endcsname\relax\def\natexlab#1{#1}\fi
\providecommand{\url}[1]{\href{#1}{#1}}
\providecommand{\doi}[1]{doi:~\href{http://doi.org/#1}{\nolinkurl{#1}}}
\providecommand{\doeprint}[1]{\href{http://ascl.net/#1}{\nolinkurl{http://ascl.net/#1}}}
\providecommand{\doarXiv}[1]{\href{https://arxiv.org/abs/#1}{\nolinkurl{https://arxiv.org/abs/#1}}}

\bibitem[{{Aasi} {et~al.}(2015)}]{LIGO+2015}
{Aasi}, J., {et~al.} 2015, Classical and Quantum Gravity, 32, 074001, \doi{10.1088/0264-9381/32/7/074001}

\bibitem[{{Abbott} {et~al.}(2017{\natexlab{a}})}]{Abbott+2017a}
{Abbott}, B.~P., {et~al.} 2017{\natexlab{a}}, \prl, 119, 161101, \doi{10.1103/PhysRevLett.119.161101}

\bibitem[{{Abbott} {et~al.}(2017{\natexlab{b}})}]{Abbott+2017b}
---. 2017{\natexlab{b}}, \apjl, 848, L12, \doi{10.3847/2041-8213/aa91c9}

\bibitem[{{Abbott} {et~al.}(2017{\natexlab{c}})}]{Abbott+2017c}
---. 2017{\natexlab{c}}, \apjl, 848, L13, \doi{10.3847/2041-8213/aa920c}

\bibitem[{{Acernese} {et~al.}(2015)}]{Virgo+2015}
{Acernese}, F., {et~al.} 2015, Classical and Quantum Gravity, 32, 024001, \doi{10.1088/0264-9381/32/2/024001}

\bibitem[{{Arnaud}(1996)}]{Arnaud+1996}
{Arnaud}, K.~A. 1996, \asp, 101, 17

\bibitem[{{Beniamini} {et~al.}(2019)}]{Beniamini+2019}
{Beniamini}, P., {et~al.} 2019, \mnras, 483, 840, \doi{10.1093/mnras/sty3093}

\bibitem[{{Bennett} {et~al.}(2014)}]{Bennett+2014}
{Bennett}, C.~L., {et~al.} 2014, \apj, 794, 135, \doi{10.1088/0004-637X/794/2/135}

\bibitem[{{Berger} {et~al.}(2007)}]{Berger+2007}
{Berger}, E., {et~al.} 2007, \apj, 664, 1000, \doi{10.1086/518762}

\bibitem[{{Berger} {et~al.}(2013)}]{Berger+2013}
---. 2013, \apj, 765, 121, \doi{10.1088/0004-637X/765/2/121}

\bibitem[{{Bromberg} {et~al.}(2018)}]{Bromberg+2018}
{Bromberg}, O., {et~al.} 2018, \mnras, 475, 2971, \doi{10.1093/mnras/stx3316}

\bibitem[{{Chornock} {et~al.}(2013)}]{Chornock+2013}
{Chornock}, R., {et~al.} 2013, GCN Circ., 15307, 1

\bibitem[{{de Ugarte Postigo} {et~al.}(2020)}]{Postigo+2020}
{de Ugarte Postigo}, A., {et~al.} 2020, GCN Circ., 29132, 1

\bibitem[{{Efron} \& {Petrosian}(1992)}]{Petrosian+1992}
{Efron}, B., \& {Petrosian}, V. 1992, \apj, 399, 345, \doi{10.1086/171931}

\bibitem[{{Eichler} {et~al.}(1989)}]{Eichler+1989}
{Eichler}, D., {et~al.} 1989, \nat, 340, 126, \doi{10.1038/340126a0}

\bibitem[{{Fong} {et~al.}(2013)}]{Fong+2013}
{Fong}, W., {et~al.} 2013, \apj, 769, 56, \doi{10.1088/0004-637X/769/1/56}

\bibitem[{{Fong} {et~al.}(2022)}]{Fong+2022}
{Fong}, W.-f., {et~al.} 2022, \apj, 940, 56, \doi{10.3847/1538-4357/ac91d0}

\bibitem[{{Goldstein} {et~al.}(2017){Goldstein}, {Veres}, {et~al.}}]{Goldstein+2017}
{Goldstein}, A., {Veres}, P., {et~al.} 2017, \apjl, 848, L14, \doi{10.3847/2041-8213/aa8f41}

\bibitem[{{Golenetskii} {et~al.}(1983)}]{Golenetskii+1983}
{Golenetskii}, S.~V., {et~al.} 1983, \nat, 306, 451, \doi{10.1038/306451a0}

\bibitem[{{Goodman}(1986)}]{Goodman1986}
{Goodman}, J. 1986, \apjl, 308, L47, \doi{10.1086/184741}

\bibitem[{{Gottlieb} {et~al.}(2018)}]{Gottlieb+2018}
{Gottlieb}, O., {et~al.} 2018, \mnras, 479, 588, \doi{10.1093/mnras/sty1462}

\bibitem[{{Granot} {et~al.}(2018)}]{Granot+2018}
{Granot}, J., {et~al.} 2018, \mnras, 481, 1597, \doi{10.1093/mnras/sty2308}

\bibitem[{{Guo} {et~al.}(2020)}]{Guo+2020}
{Guo}, Q., {et~al.} 2020, \apj, 896, 83, \doi{10.3847/1538-4357/ab8f9d}

\bibitem[{{Hamidani} \& {Ioka}(2023{\natexlab{a}})}]{Hamidai+2023a}
{Hamidani}, H., \& {Ioka}, K. 2023{\natexlab{a}}, \mnras, 520, 1111, \doi{10.1093/mnras/stad041}

\bibitem[{{Hamidani} \& {Ioka}(2023{\natexlab{b}})}]{Hamidani+2023b}
---. 2023{\natexlab{b}}, \mnras, 524, 4841, \doi{10.1093/mnras/stad1933}

\bibitem[{{Hamidani} {et~al.}(2020)}]{Hamidani+2020}
{Hamidani}, H., {et~al.} 2020, \mnras, 491, 3192, \doi{10.1093/mnras/stz3231}

\bibitem[{{Hjorth} {et~al.}(2003)}]{Hjorth+2003}
{Hjorth}, J., {et~al.} 2003, \nat, 423, 847, \doi{10.1038/nature01750}

\bibitem[{{Hjorth} {et~al.}(2005)}]{Hjorth+2005}
---. 2005, \nat, 437, 859, \doi{10.1038/nature04174}

\bibitem[{{Ioka} \& {Nakamura}(2018)}]{Ioka+2018}
{Ioka}, K., \& {Nakamura}, T. 2018, Progress of Theoretical and Experimental Physics, 2018, 043E02, \doi{10.1093/ptep/pty036}

\bibitem[{{Ioka} \& {Nakamura}(2019)}]{Ioka+2019}
---. 2019, \mnras, 487, 4884, \doi{10.1093/mnras/stz1650}

\bibitem[{{Ito} {et~al.}(2023)}]{Ito+2023}
{Ito}, H., {et~al.} 2023, arXiv e-prints, arXiv:2307.10023, \doi{10.48550/arXiv.2307.10023}

\bibitem[{{Iwamoto} {et~al.}(1998)}]{Iwamoto+1998}
{Iwamoto}, K., {et~al.} 1998, \nat, 395, 672, \doi{10.1038/27155}

\bibitem[{{Kasliwal} {et~al.}(2017)}]{Kasliwal+2017}
{Kasliwal}, M.~M., {et~al.} 2017, Science, 358, 1559, \doi{10.1126/science.aap9455}

\bibitem[{{Kisaka} {et~al.}(2017)}]{Kisaka+2017}
{Kisaka}, S., {et~al.} 2017, \apj, 846, 142, \doi{10.3847/1538-4357/aa8775}

\bibitem[{{Kouveliotou} {et~al.}(1993)}]{Kouveliotou+1993}
{Kouveliotou}, C., {et~al.} 1993, \apjl, 413, L101, \doi{10.1086/186969}

\bibitem[{{Lazzati} {et~al.}(2017)}]{Lazzati+2017}
{Lazzati}, D., {et~al.} 2017, \apjl, 848, L6, \doi{10.3847/2041-8213/aa8f3d}

\bibitem[{{Lazzati} {et~al.}(2018)}]{Lazzati+2018}
---. 2018, \prl, 120, 241103, \doi{10.1103/PhysRevLett.120.241103}

\bibitem[{{Levan} {et~al.}(2015)}]{Levan+2015}
{Levan}, A., {et~al.} 2015, The Astronomer's Telegram, 6873, 1

\bibitem[{{Levan} {et~al.}(2016)}]{Levan+2016}
{Levan}, A.~J., {et~al.} 2016, GCN Circ., 19846, 1

\bibitem[{{Levan} {et~al.}(2017)}]{Levan+2017}
---. 2017, \apjl, 848, L28, \doi{10.3847/2041-8213/aa905f}

\bibitem[{{Lien} {et~al.}(2016)}]{Lien+2016}
{Lien}, A., {et~al.} 2016, \apj, 829, 7, \doi{10.3847/0004-637X/829/1/7}

\bibitem[{{Lloyd-Ronning} {et~al.}(2020)}]{Ronning+2020}
{Lloyd-Ronning}, N., {et~al.} 2020, \mnras, 494, 4371, \doi{10.1093/mnras/staa1057}

\bibitem[{{Lloyd-Ronning} {et~al.}(2002)}]{Lloyd+2002}
{Lloyd-Ronning}, N.~M., {et~al.} 2002, \apj, 574, 554, \doi{10.1086/341059}

\bibitem[{{Lloyd-Ronning} {et~al.}(2019)}]{Ronning+2019}
---. 2019, \mnras, 488, 5823, \doi{10.1093/mnras/stz2155}

\bibitem[{{L{\'o}pez-C{\'a}mara} {et~al.}(2014)}]{Lopez+2014}
{L{\'o}pez-C{\'a}mara}, D., {et~al.} 2014, \mnras, 442, 2202, \doi{10.1093/mnras/stu1016}

\bibitem[{{Lu} {et~al.}(2012)}]{Lu+2012}
{Lu}, R.-J., {et~al.} 2012, \apj, 756, 112, \doi{10.1088/0004-637X/756/2/112}

\bibitem[{{MacFadyen} \& {Woosley}(1999)}]{MacFadyen+1999}
{MacFadyen}, A.~I., \& {Woosley}, S.~E. 1999, \apj, 524, 262, \doi{10.1086/307790}

\bibitem[{{Mandhai} {et~al.}(2018)}]{Mandhai+2018}
{Mandhai}, S., {et~al.} 2018, Galaxies, 6, 130, \doi{10.3390/galaxies6040130}

\bibitem[{Markwardt {et~al.}(2007)}]{Markwardt+2007}
Markwardt, C., {et~al.} 2007, NASA/GSFC, Greenbelt, MD, 6

\bibitem[{{Matsumoto} \& {Piran}(2020)}]{Matsumoto+2020}
{Matsumoto}, T., \& {Piran}, T. 2020, \mnras, 492, 4283, \doi{10.1093/mnras/staa050}

\bibitem[{{McEnery} {et~al.}(2012)}]{McEnery+2012}
{McEnery}, J.~E., {et~al.} 2012, Optical Engineering, 51, 011012, \doi{10.1117/1.OE.51.1.011012}

\bibitem[{{Mooley} {et~al.}(2018)}]{Mooley+2018}
{Mooley}, K.~P., {et~al.} 2018, \nat, 561, 355, \doi{10.1038/s41586-018-0486-3}

\bibitem[{{Murguia-Berthier} {et~al.}(2017)}]{Murguia+2017}
{Murguia-Berthier}, A., {et~al.} 2017, \apjl, 848, L34, \doi{10.3847/2041-8213/aa91b3}

\bibitem[{{Nakar} {et~al.}(2018)}]{Nakar+2018}
{Nakar}, E., {et~al.} 2018, \apj, 867, 18, \doi{10.3847/1538-4357/aae205}

\bibitem[{{Paczynski}(1986)}]{Paczynski1986}
{Paczynski}, B. 1986, \apjl, 308, L43, \doi{10.1086/184740}

\bibitem[{{Parsotan} {et~al.}(2018)}]{Parsotan+2018}
{Parsotan}, T., {et~al.} 2018, \apj, 869, 103, \doi{10.3847/1538-4357/aaeed1}

\bibitem[{{Paul}(2018)}]{Paul+2018}
{Paul}, D. 2018, \mnras, 477, 4275, \doi{10.1093/mnras/sty840}

\bibitem[{{Price-Whelan} {et~al.}(2022)}]{Astropy2022}
{Price-Whelan}, A.~M., {et~al.} 2022, \apj, 935, 167, \doi{10.3847/1538-4357/ac7c74}

\bibitem[{{Rhoads}(1999)}]{Rhoads1999}
{Rhoads}, J.~E. 1999, \apj, 525, 737, \doi{10.1086/307907}

\bibitem[{{Roming} {et~al.}(2006)}]{Roming+2006}
{Roming}, P. W.~A., {et~al.} 2006, \apj, 651, 985, \doi{10.1086/508054}

\bibitem[{{Rossi} {et~al.}(2021)}]{Rossi+2021}
{Rossi}, A., {et~al.} 2021, GCN Circ., 31107, 1

\bibitem[{{Rouco Escorial} {et~al.}(2023)}]{Escorial+2023}
{Rouco Escorial}, A., {et~al.} 2023, \apj, 959, 13, \doi{10.3847/1538-4357/acf830}

\bibitem[{{Sakamoto} {et~al.}(2011)}]{Sakamoto+2011}
{Sakamoto}, T., {et~al.} 2011, \apjs, 195, 2, \doi{10.1088/0067-0049/195/1/2}

\bibitem[{{Sari} {et~al.}(1999)}]{Sari+1999}
{Sari}, R., {et~al.} 1999, \apjl, 519, L17, \doi{10.1086/312109}

\bibitem[{{Savchenko} {et~al.}(2017)}]{Savchenko+2017}
{Savchenko}, V., {et~al.} 2017, \apjl, 848, L15, \doi{10.3847/2041-8213/aa8f94}

\bibitem[{{Scargle}(1998)}]{Scargle1998}
{Scargle}, J.~D. 1998, \apj, 504, 405, \doi{10.1086/306064}

\bibitem[{{Scargle} {et~al.}(2013)}]{Scargle+2013}
{Scargle}, J.~D., {et~al.} 2013, \apj, 764, 167, \doi{10.1088/0004-637X/764/2/167}

\bibitem[{{Thoene} {et~al.}(2010)}]{Thoene+2010}
{Thoene}, C.~C., {et~al.} 2010, GCN Circ., 10971, 1

\bibitem[{{Troja} {et~al.}(2016)}]{Troja+2016}
{Troja}, E., {et~al.} 2016, \apjl, 822, L8, \doi{10.3847/2041-8205/822/1/L8}

\bibitem[{{Troja} {et~al.}(2019)}]{Troja+2019}
---. 2019, \mnras, 489, 1919, \doi{10.1093/mnras/stz2248}

\bibitem[{{Veres} {et~al.}(2019)}]{Veres+2019}
{Veres}, P., {et~al.} 2019, \apj, 882, 53, \doi{10.3847/1538-4357/ab31aa}

\bibitem[{{von Kienlin} {et~al.}(2019)}]{Kienlin+2019}
{von Kienlin}, A., {et~al.} 2019, \apj, 876, 89, \doi{10.3847/1538-4357/ab10d8}

\bibitem[{{von Kienlin} {et~al.}(2020)}]{Kienlin+2020}
---. 2020, \apj, 893, 46, \doi{10.3847/1538-4357/ab7a18}

\bibitem[{{Winkler} {et~al.}(2003)}]{Winkler+2003}
{Winkler}, C., {et~al.} 2003, \aap, 411, L1, \doi{10.1051/0004-6361:20031288}

\bibitem[{{Woosley}(1993)}]{Woosley1993}
{Woosley}, S.~E. 1993, \apj, 405, 273, \doi{10.1086/172359}

\bibitem[{{Yonetoku} {et~al.}(2004)}]{Yonetoku+2004}
{Yonetoku}, D., {et~al.} 2004, \apj, 609, 935, \doi{10.1086/421285}

\bibitem[{{Yonetoku} {et~al.}(2005)}]{Yonetoku+2005}
---. 2005, \mnras, 362, 1114, \doi{10.1111/j.1365-2966.2005.09398.x}

\bibitem[{{Yonetoku} {et~al.}(2010)}]{Yonetoku+2010}
---. 2010, \pasj, 62, 1495, \doi{10.1093/pasj/62.6.1495}

\bibitem[{{Yonetoku} {et~al.}(2014)}]{Yonetoku+2014}
---. 2014, \apj, 789, 65, \doi{10.1088/0004-637X/789/1/65}

\bibitem[{{Zhang} \& {Wang}(2018)}]{Zhang+2018}
{Zhang}, G.~Q., \& {Wang}, F.~Y. 2018, \apj, 852, 1, \doi{10.3847/1538-4357/aa9ce5}

\bibitem[{{Zhang} {et~al.}(2012)}]{Zhang+2012}
{Zhang}, Z.~B., {et~al.} 2012, \apj, 755, 55, \doi{10.1088/0004-637X/755/1/55}

\end{thebibliography}
\bibliographystyle{aasjournal}

\end{document}